\documentclass[aps,onecolumn,10pt]{revtex4}
\usepackage{amsmath}
\usepackage{amssymb}
\usepackage{hyperref}
\hypersetup{colorlinks=true} 
\numberwithin{equation}{section}
\numberwithin{equation}{section}

\hypersetup{colorlinks=true} 
\begin{document}
\allowdisplaybreaks
\setcounter{equation}{0}

\title{Cosmological Fluctuations on the Light Cone}

\author{Asanka Amarasinghe  and Philip D. Mannheim}
\affiliation{Department of Physics, University of Connecticut, Storrs, CT 06269, USA \\
asanka.amarasinghe@uconn.edu,philip.mannheim@uconn.edu\\ }

\date{March 29 2021}

\begin{abstract}

In studying temperature fluctuations in the cosmic microwave background Weinberg has noted that some ease of calculation and insight can be achieved by looking at the structure of the perturbed light cone on which the perturbed photons propagate. In his approach Weinberg worked in a specific gauge and specialized to fluctuations around the standard Robertson-Walker cosmological model with vanishing spatial three-curvature. In this paper we generalize this analysis by providing a gauge invariant treatment in which no choice of gauge is made, and by considering geometries with non-vanishing spatial three-curvature. By using the scalar, vector, tensor fluctuation basis we find that the relevant gauge invariant combinations that appear in the light cone temperature fluctuations have no explicit dependence on the spatial curvature even if the spatial curvature of the background geometry is nonvanishing. We find that a not previously considered, albeit not too consequential,  temperature fluctuation at the observer has to be included in order to enforce gauge invariance. As well as working with comoving time we also work with conformal time in which a background metric of any given spatial three-curvature can be written as a time-dependent conformal factor (the comoving time expansion radius as written in conformal time) times a static Robertson-Walker geometry of the same spatial three-curvature.  For temperature fluctuations on the light cone this conformal factor drops out identically. Thus the gauge invariant combinations that appear in the photon temperature fluctuations have no explicit dependence on either the conformal factor or the spatial three-curvature at all.

\end{abstract}

\maketitle

\section{Introduction}
\label{S1}

In analyzing the cosmological perturbations that can be measured in the cosmic microwave background (CMB)  it is very convenient to use the scalar, vector, tensor (SVT) basis for the fluctuations as developed in  \cite{Lifshitz1946}  and \cite{Bardeen1980}. In this basis the fluctuations are characterized according to how they transform under three-dimensional spatial rotations, and in this form the basis  has been applied extensively in cosmological perturbation theory (see e.g. \cite{Kodama1984,Mukhanov1992,Stewart1990,Ma1995,Bertschinger1996,Zaldarriaga1998} and \cite{Dodelson2003,Mukhanov2005,Weinberg2008,Lyth2009,Ellis2012}). With the SVT expansion being based on quantities that transform as three-dimensional scalars, vectors and tensors, as such it is particularly well suited to Robertson-Walker geometries because such geometries have a spatial sector that is maximally three-symmetric. While not manifestly covariant, the scalar, vector, tensor expansion is covariant as it leads to equations that involve appropriate combinations of the scalars, vectors and tensors that are fully four-dimensionally (i.e., not just three-dimensionally) gauge invariant, this being all that one needs  for covariance. Given that the fluctuation equations are gauge invariant one can of course work in any particular gauge that might be convenient. However, it is also informative to use a formalism that is manifestly fully gauge invariant throughout and that does not involve any specific choice of gauge at all \cite{footnoteA}. Such an approach to cosmological perturbation theory has been followed in \cite{Amarasinghe2018}, \cite{Phelps2019}, and \cite{Mannheim2020}, and in this paper we apply this approach to light-cone fluctuations in the CMB \cite{footnoteB}. As we show, the relevant gauge invariant combinations that appear in the temperature fluctuations have no explicit dependence on the spatial curvature even if the spatial curvature of the background geometry is nonvanishing. This result meshes well with the observed CMB temperature fluctuations since in standard gravity they are found to favor vanishing spatial three-curvature. As well as work in comoving time we also work in conformal time where the background metric can be written as an overall time-dependent conformal factor times a static metric, and find that the photon temperature fluctuations are completely independent of the conformal factor. Thus, quite strikingly, light cone temperature fluctuations have no explicit dependence on either the conformal factor or the spatial three-curvature at all.

Our interest here is in fluctuations around cosmological backgrounds that are described by Robertson-Walker metrics of the form
\begin{eqnarray}
ds^2 &=& -g_{\mu\nu}dx^{\mu}dx^{\nu}= dt^2-a^2(t)\left[\frac{dr^2}{1-kr^2} +r^2d\theta^2+r^2\sin^2\theta d\phi^2\right]=dt^2-a^2(t)\tilde{\gamma}_{ij}dx^idx^j,
\label{1.1e}
\end{eqnarray}
where $\tilde{\gamma}_{ij}$ is the metric associated with the spatial three-space, and where we use the conventions given in \cite{Weinberg1972} in which the signature of the metric is $(-1,1,1,1)$. For the fluctuations the fluctuation metric can be written as 
\begin{align}
ds^2&=-h_{\mu\nu}dx^{\mu}dx^{\nu}=\left[2\phi dt^2 -2a(t)(\tilde{\nabla}_i B +B_i)dt dx^i - a^2(t)[-2\psi\tilde{\gamma}_{ij} +2\tilde{\nabla}_i\tilde{\nabla}_j E + \tilde{\nabla}_i E_j + \tilde{\nabla}_j E_i + 2E_{ij}]dx^i dx^j\right].
\label{1.2e}
\end{align}
In (\ref{1.2e})  $\tilde{\nabla}_i=\partial/\partial x^i$ and  $\tilde{\nabla}^i=\tilde{\gamma}^{ij}\tilde{\nabla}_j$  (with Latin indices) are defined with respect to the background three-space metric $\tilde{\gamma}_{ij}$, and $(1,2,3)$ denote $(r,\theta,\phi)$. And with
\begin{eqnarray}
\tilde{\gamma}^{ij}\tilde{\nabla}_j V_i=\tilde{\gamma}^{ij}[\partial_j V_i-\tilde{\Gamma}^{k}_{ij}V_k]
\label{1.3e}
\end{eqnarray}
for any three-vector $V_i$ in a three-space with three-space connection $\tilde{\Gamma}^{k}_{ij}$, the elements of (\ref{1.2e}) are required to obey
\begin{eqnarray}
\tilde{\gamma}^{ij}\tilde{\nabla}_j B_i = 0,\quad \tilde{\gamma}^{ij}\tilde{\nabla}_j E_i = 0, \quad E_{ij}=E_{ji},\quad \tilde{\gamma}^{jk}\tilde{\nabla}_kE_{ij} = 0, \quad\tilde{\gamma}^{ij}E_{ij} = 0.
\label{1.4e}
\end{eqnarray}
With the  three-space sector of the background geometry being maximally three-symmetric, it is described by a Riemann tensor of the form
\begin{eqnarray}
\tilde{R}_{ijk\ell}=k[\tilde{\gamma}_{jk}\tilde{\gamma}_{i\ell}-\tilde{\gamma}_{ik}\tilde{\gamma}_{j\ell}].
\label{1.5e}
\end{eqnarray}

As written, (\ref{1.2e}) contains ten elements, whose transformations are defined with respect to the background spatial sector as four three-dimensional scalars ($\phi$, $B$, $\psi$, $E$) each with one degree of freedom, two transverse three-dimensional vectors ($B_i$, $E_i$) each with two independent degrees of freedom, and one symmetric three-dimensional transverse-traceless tensor ($E_{ij}$) with two degrees of freedom. The great utility of this basis is that since the cosmological fluctuation equations are gauge invariant, only gauge-invariant scalar, vector, or tensor combinations of the components of the scalar, vector, tensor basis can appear in the fluctuation equations. By studying gauge transformations of the form $h_{\mu\nu}\rightarrow h_{\mu\nu}-\nabla_{\mu}\epsilon_{\nu}-\nabla_{\nu}\epsilon_{\mu}$
in \cite{Amarasinghe2018} it was shown in the $k=0$ case that for the fluctuations associated with the metric given in (\ref{1.1e}) and (\ref{1.2e}) and with $a(t)$ being an arbitrary function of $t$, the gauge-invariant metric combinations are \cite{footnoteC}
\begin{align}
\alpha=\phi + \psi + a\frac{\partial B}{\partial t} - a^2\frac{ \partial^2E}{\partial t^2}-a\frac{da}{dt} \frac{\partial E}{\partial t},\quad  \gamma= - \left(\frac{da }{dt}\right)^{-1}\psi + B - a\frac{\partial E}{\partial t},\quad  B_i-a\frac{\partial E_i}{ \partial t},  \quad E_{ij},
\label{1.6e}
\end{align}
for a total of six degrees of freedom, just as required since one can make four coordinate transformations  on the initial ten fluctuation components. As we shall see below, the light-cone fluctuations will explicitly depend on these specific combinations, together with combinations of these quantities and some quantities associated with the perturbed energy-momentum tensor. Also, by further study of the $h_{\mu\nu}\rightarrow h_{\mu\nu}-\nabla_{\mu}\epsilon_{\nu}-\nabla_{\nu}\epsilon_{\mu}$ gauge transformation we will extend (\ref{1.6e}) to the non-zero $k$ case, and show that these gauge invariant combinations continue to hold  unaltered without acquiring any explicit dependence on $k$ at all. 

In the following we shall also have occasion to work in conformal time where 
\begin{eqnarray}
\tau=\int \frac{dt}{a(t)}.
\label{1.7e}
\end{eqnarray}
On setting $\Omega(\tau)=a(t)$, (\ref{1.1e}), (\ref{1.2e}) and (\ref{1.6e}) take the somewhat more compact form \cite{footnoteD}
\begin{align}
ds^2&=\Omega^2(\tau)\left[d\tau^2-\frac{dr^2}{1-kr^2}-r^2d\theta^2-r^2\sin^2\theta d\phi^2\right]
\nonumber\\
&+\Omega^2(\tau)\left[2\phi d\tau^2 -2(\tilde{\nabla}_i B +B_i)d\tau dx^i - [-2\psi\tilde{\gamma}_{ij} +2\tilde{\nabla}_i\tilde{\nabla}_j E + \tilde{\nabla}_i E_j + \tilde{\nabla}_j E_i + 2E_{ij}]dx^i dx^j\right],
\label{1.8e}
\end{align}
and
\begin{align}
\alpha=\phi + \psi + \frac{\partial B}{\partial\tau} - \frac{\partial^2E}{\partial\tau^2},\quad  \gamma= - \left(\frac{d \Omega}{d \tau}\right)^{-1}\Omega \psi + B - \frac{\partial E}{\partial\tau},\quad  B_i-\frac{\partial E_i}{\partial \tau},  \quad E_{ij}.
\label{1.9e}
\end{align}

In a study in \cite{Weinberg2008} Weinberg considered the propagation of CMB photon fluctuations by setting $ds^2=0$ in (\ref{1.1e}) and (\ref{1.2e}). With the primary interest in cosmology being in geometries with $k=0$ \cite{Guth1981,Bahcall2000,deBernardis2000,Tegmark2004}, on choosing a gauge and on dropping the vector sector that is dynamically suppressed  in the standard cosmological model, Weinberg studied radial modes with $ds^2=0$ in the geometry
\begin{eqnarray}
ds^2 &=& dt^2-a^2(t)\left[dr^2 +r^2d\theta^2+r^2\sin^2\theta d\phi^2\right]
+2\phi dt^2 - a^2(t)[-2\psi\tilde{\gamma}_{ij} +2\tilde{\nabla}_i\tilde{\nabla}_j  E  + 2E_{ij}]dx^i dx^j.
\label{1.10e}
\end{eqnarray}
With CMB photons propagating on radial trajectories in the direction $\hat{n}$ that originate at last scattering at $r_L$, $t_L$ and reach an observer at $r=0$ at the current time $t_0$, and with $\tilde{\nabla}_r\tilde{\nabla}_r  E=\partial_r\partial_r E-\tilde{\Gamma}^k_{rr}\tilde{\nabla}_kE=\partial_r\partial_r E$ in flat three-space polar coordinates, the radial modes travel on
\begin{eqnarray}
 dt^2-a^2(t)dr^2 +2\phi dt^2 - a^2(t)[-2\psi+2\partial_r\partial_r E  + 2E_{rr}]=0.
\label{1.11e}
\end{eqnarray}
For such modes the temperature fluctuation is generically given by 
\begin{align}
\left(\frac{\Delta T(\hat{n})}{T_0}\right)=\left(\frac{\Delta T(\hat{n})}{T_0}\right)_{\rm early}+\left(\frac{\Delta T(\hat{n})}{T_0}\right)_{\rm ISW}+\left(\frac{\Delta T(\hat{n})}{T_0}\right)_{\rm late},
\label{1.12e}
\end{align}
where 'early' denotes the early time $t_L$, 'late' denotes the current or late time $t_0$, and ISW denotes the integrated Sachs-Wolfe effect \cite{Sachs1967} as integrated from $t_L$ to $t_0$. With $T_0$ denoting the current temperature and $\bar{T}_L$ denoting the temperature at last scattering, these quantities evaluate to \cite{Weinberg2008} 
\begin{align}
\left(\frac{\Delta T(\hat{n})}{T_0}\right)_{\rm early}=&\left(\phi-a^2\frac{\partial^2 E}{\partial t^2}-a\frac{da}{dt}\frac{\partial E}{\partial t}-a\frac{\partial^2 E}{\partial t\partial r}\right)_{r=r_L,t=t_L}+\frac{\delta T(r_L\hat{n},t_L)}{\bar{T}(t_L)}
-\frac{1}{a(t_L)}\left( \frac{\partial}{\partial r}\delta u(r\hat{n},t_L)\right)_{r=r_L},
\label{1.13e}
\end{align}
\begin{align}
\left(\frac{\Delta T(\hat{n})}{T_0}\right)_{\rm ISW}=&\int_{t_L}^{t_0}dt\left[\frac{\partial}{\partial t}\left(\phi+\psi-E_{rr}-a^2\frac{\partial^2E}{\partial t^2}-a\frac{da}{dt}\frac{\partial E}{\partial t}\right)\right]_{r=s(t)},
\label{1.14e}
\end{align}
\begin{align}
\left(\frac{\Delta T(\hat{n})}{T_0}\right)_{\rm late}=&-\left(\phi-a^2\frac{\partial^2 E}{\partial t^2}-a\frac{\partial a}{\partial t}\frac{\partial E}{\partial t}-a\frac{\partial^2 E}{\partial t\partial r}\right)_{r=0,t=t_0}+\frac{1}{a(t_0)}\left( \frac{\partial}{\partial r}\delta u(r\hat{n},t_0)\right)_{r=0},
\label{1.15e}
\end{align}
with $a(t)$ being arbitrary. Here $s(t)$ is the background radial trajectory solution to $dr/dt=-1/a(t)$ \cite{footnoteE} as chosen to obey $s(t_L)=r_L$, and the $\delta u(r\hat{n},t)_{r_L}$ and $\delta u(r\hat{n},t)_{r=0}$ terms are due to the change with time of the radial coordinates $r_L$ and $r_0$ of the points where the light signals are being emitted and detected \cite{Weinberg2008}. We now generalize this result, and note that since no use of any gravitational fluctuation equation is made the result is a purely kinematic one that holds in any cosmology in any covariant, metric-based theory of gravity.

\section{General Photon Fluctuations in Robertson-Walker Geometries with $k=0$}
\label{S2}

We consider photon fluctuations around a $k=0$ geometry with line element
\begin{align}
ds^2 &= dt^2-a^2(t)\left[dr^2 +r^2d\theta^2+r^2\sin^2\theta d\phi^2\right]
\nonumber\\
&+\left[2\phi dt^2 -2a(t)(\tilde{\nabla}_i B +B_i)dt dx^i - a^2(t)[-2\psi\tilde{\gamma}_{ij} +2\tilde{\nabla}_i\tilde{\nabla}_j E + \tilde{\nabla}_i E_j + \tilde{\nabla}_j E_i + 2E_{ij}]dx^i dx^j\right],
\label{2.1e}
\end{align}
where $a(t)$ is  arbitrary. In this geometry radial modes with fixed $\theta$ and $\phi$ obey 
\begin{align}
&dt^2-a^2(t)dr^2+2\phi dt^2 -2a(t)(\partial_rB +B_r)dt dr - a^2(t)[-2\psi +2\partial_r\partial_rE + 2\partial_rE_r + 2E_{rr}]dr^2=0.
\label{2.2e}
\end{align}
To simplify the writing we set
\begin{align}
\frac{h_{rr}}{a^2(t)}=f_{rr}&=-2\psi +2\partial_r\partial_rE + 2\partial_rE_r+ 2E_{rr},
\label{2.3e}
\end{align}
with solutions to (\ref{2.2e}) then obeying
\begin{align}
\frac{dr}{dt}=&\frac{-a(t)(\partial_rB+B_r)\pm a(t)[(\partial_rB+B_r)^2+(1+f_{rr})(1+2\phi)]^{1/2}}{a^2(t)(1+f_{rr})}.
\label{2.4e}
\end{align}
On taking the negative square root, to lowest non-trivial order (\ref{2.4e}) takes the form
\begin{align}
\frac{dr}{dt}=&-\frac{1}{a(t)}+\frac{1}{a(t)}\left[-\phi+\frac{f_{rr}}{2}-\partial_rB-B_r\right],
\label{2.5e}
\end{align}
and reduces to $dr/dt=-1/a$ in the absence of the perturbation. On setting
\begin{align}
N[r(t)\hat{n},t]=-\phi-\partial_rB-B_r-\psi +\frac{\partial^2 E}{\partial r^2} + \frac{\partial E_r}{\partial r} + E_{rr},
\label{2.6e}
\end{align}
solutions to (\ref{2.5e}) take the form \cite{footnoteF}
\begin{align}
r(t)=r_L-\int_{t_L}^{t}\frac{dt}{a(t)}+\int_{t_L}^{t}\frac{dt^{\prime}}{a(t^{\prime})}N[s(t')\hat{n},t'],
\label{2.7e}
\end{align}
where $s(t)$ is the solution to (\ref{2.7e}) when $N[r(t)\hat{n},t]$ is absent, viz.
\begin{align}
s(t)=r_L-\int_{t_L}^{t}\frac{dt}{a(t)},
\label{2.8e}
\end{align}
as normalized so that $s(t_L)=r_L$. We note that $s(t_0)$ is given by
\begin{align}
s(t_0)=r_L-\int_{t_L}^{t_0}\frac{dt}{a(t)},
\label{2.9e}
\end{align}
while at $t=t_0$ (\ref{2.7e}) yields  
\begin{align}
r(t_0)=r_0=r_L-\int_{t_L}^{t_0}\frac{dt}{a(t)}+\int_{t_L}^{t_0}\frac{dt}{a(t)}N[s(t)\hat{n},t].
\label{2.10e}
\end{align}
Since we set $r(t_0)=r_0=0$ it follows that $s(t_0)$ is given by
\begin{align}
s(t_0)=-\int_{t_L}^{t_0}\frac{dt}{a(t)}N[s(t)\hat{n},t],
\label{2.11e}
\end{align}
and is not zero. Thus in the absence of any perturbation a light ray that would have set off at the same $r_L$ and $t_L$ as the perturbed light ray would not reach $r=0$ at the same time as the perturbed light ray does, and it is the $s(t)$ given by (\ref{2.8e}) that should appear in $N[s(t)\hat{n},t]$. Finally, with $r(t_0)=0$ we obtain
\begin{align}
r_L+\int_{t_L}^{t_0}\frac{dt}{a(t)} \left[N[s(t)\hat{n},t]-1\right]=0.
\label{2.12e}
\end{align}

Following \cite{Weinberg2008} we now compare light wave crests leaving at last scattering at $(r_L,t_L)$ with light crests leaving at $(r_L,t_L+\delta t_L)$ at a time $\delta t_L$ later. These crests respectively arrive  at the observer at times $t_0$ and $t_0+\delta t_0$. As well as this change in $r(t)$ there is also a change in the $s(t)$ that appears in $N[s(t)\hat{n},t]$ as the relevant $s(t)$ is now the one that leaves at $t_L+\delta t_L$ \cite{footnoteG}. According to (\ref{2.8e}) this change is given by 
\begin{align}
\delta s(t)=-\int_{t_L+\delta t_L}^{t}\frac{dt}{a(t)}+\int_{t_L}^{t}\frac{dt}{a(t)}=\frac{\delta t_L}{a(t_L)},
\label{2.13e}
\end{align}
with the associated change in $N[s(t)\hat{n},t]$ thus being given by
\begin{align}
\delta N[s(t)\hat{n},t]=\left(\frac{\partial N[r\hat{n},t]}{\partial r}\right)_{r=s(t)}\frac{\delta t_L}{a(t_L)}.
\label{2.14e}
\end{align}
Thus via variation of the endpoints and integrand of the two integrals in (\ref{2.10e}) while holding $r_L$ and $r_0$ fixed we obtain
\begin{align}
-\frac{\delta t_L}{a(t_L)}\left[N[r(t_L)\hat{n},t_L]-1\right]+\frac{\delta t_L}{a(t_L)}\int_{t_L}^{t_0}\frac{dt}{a(t)}\left(\frac{\partial N[r\hat{n},t]}{\partial r}\right)_{r=s(t)}+\frac{\delta t_0}{a(t_0)}\left[N[0,t_0]-1\right]=0.
\label{2.15e}
\end{align}
However, as well as a fluctuation in the velocity of the radial photon mode there is also fluctuation in the velocity $u^r$ of the overall background perfect fluid of which both the radial photon modes and the charged particles that emit or detect them are a part, a point we shall comment on further below. Such fluctuations occur at both $r_L$ and $r_0$, with a light signal emitted at $t_L+\delta t_L$ being emitted not from $r_L$ but from $r_L+\delta r_L$, and a light signal detected at $t_0+\delta t_0$ being detected not at $r_0$ but at $r_0+\delta r_0$, where $\delta r_L=\delta t_L\delta u^r(r_L\hat{n},t_L)$ and $\delta r_0=\delta t_0\delta u^r(0,t_0)$. We thus modify the variation of (\ref{2.10e}) to 
\begin{align}
\delta t_L\delta u^r(r_L\hat{n},t_L)&-\frac{\delta t_L}{a(t_L)}\left[N[r(t_L)\hat{n},t_L]-1\right]+\frac{\delta t_L}{a(t_L)}\int_{t_L}^{t_0}\frac{dt}{a(t)}\left(\frac{\partial N[r\hat{n},t]}{\partial r}\right)_{r=s(t)}
\nonumber\\
-\delta t_0\delta u^r(0,t_0)&+\frac{\delta t_0}{a(t_0)}\left[N[0,t_0]-1\right]=0.
\label{2.16f}
\end{align}

To facilitate the evaluation of (\ref{2.16f}) we note that with $dr/dt=-1/a(t)$, to lowest non-trivial order  the total time derivative of $N[s(t)\hat{n},t]$ is given by 
\begin{align}
\frac{d}{dt}N[s(t)\hat{n},t]=\left(\frac{\partial}{\partial t}N[r\hat{n},t]\right)_{r=s(t)}-\frac{1}{a(t)}\left(\frac{\partial}{\partial r}N[r\hat{n},t]\right)_{r=s(t)}.
\label{2.17f}
\end{align}
Consequently, we can rewrite (\ref{2.16f}) as
\begin{align}
\delta t_L\delta u^r(r_L\hat{n},t_L)&-\frac{\delta t_L}{a(t_L)}\left[N[s(t_0)\hat{n},t_0]-1\right]+\frac{\delta t_L}{a(t_L)}\int_{t_L}^{t_0}dt\left(\frac{\partial N[r\hat{n},t]}{\partial t}\right)_{r=s(t)}
\nonumber\\
-\delta t_0\delta u^r(0,t_0)
&+\frac{\delta t_0}{a(t_0)}\left[N[0,t_0]-1\right]=0.
\label{2.18f}
\end{align}
While $s(t_0)$ is not zero, we note from (\ref{2.11e}) that it is of order the perturbation. Thus to first order in the perturbation we can set $s(t_0)$ equal to zero in $N[s(t_0)\hat{n},t_0]$, with (\ref{2.18f}) then taking the form 
\begin{align}
\delta t_L\delta u^r(r_L\hat{n},t_L)&-\frac{\delta t_L}{a(t_L)}\left[N[0,t_0]-1\right]+\frac{\delta t_L}{a(t_L)}\int_{t_L}^{t_0}dt\left(\frac{\partial N[r\hat{n},t]}{\partial t}\right)_{r=s(t)}
\nonumber\\
&-\delta t_0\delta u^r(0,t_0)
+\frac{\delta t_0}{a(t_0)}\left[N[0,t_0]-1\right]=0.
\label{2.19f}
\end{align}

To determine the frequency shift from $\nu_L$ at $t_L$ to $\nu_0$ at $t_0$ we need to incorporate the change in the zero-zero component of the metric as it also modulates the time behavior, so that $\nu_0/\nu_L$ is not given by the  time ratio change $\delta t_L/\delta t_0$ but by the proper time ratio change
\begin{align}
\frac{\nu_0}{\nu_L}&=\frac{(1+2\phi(r_L,t_L))^{1/2}\delta t_L}{(1+2\phi(0,t_0))^{1/2}\delta t_0}
\nonumber\\
&
=\frac{a(t_L)}{a(t_0)}\left[1+\phi(r_L,t_L)-\phi(0,t_0)-\int_{t_L}^{t_0}dt\left(\frac{\partial N[r\hat{n},t]}{\partial t}\right)_{r=s(t)}
-a(t_L)\delta u^r(r_L\hat{n},t_L)+a(t_0)\delta u^r(0,t_0)\right].
\label{2.20f}
\end{align}
For a black body propagating in a background expanding Robertson-Walker universe there is no change in the background $h\nu/kT$. However with a  temperature perturbation, in the direction $\hat{n}$ one has
\begin{align}
T(\hat{n},t_0)=\frac{\nu_0}{\nu_L}[\bar{T}(t_L)+\delta T(r_L\hat{n},t_L)],
\label{2.21f}
\end{align}
where $\delta T(r_L\hat{n},t_L)$ is the intrinsic change in the temperature at last scattering due to dynamical effects in the CMB that change the black-body energy density. Specifically, with a black body having an energy density $\rho$ given by $\rho=a T^4$, the intrinsic temperature change is  given by
\begin{align}
\frac{\delta T(r_L\hat{n},t_L)}{T(r_L\hat{n},t_L)}=\frac{\delta \rho}{4\rho},
\label{2.22f}
\end{align}
with $\delta \rho/\rho$ being determined by the dynamics. With a background perfect photon fluid with energy density $\rho$ and pressure $p$ obeying $\rho=3p$ so that $\rho$  behaves as $\rho\sim 1/a^4(t)$, the temperature change that would be observed in the absence of any perturbation is given by the isotropic and adiabatic
\begin{align}
\frac{T_0}{\bar{T}_L}=\frac{a(t_L)}{a(t_0)}=\frac{\nu_0}{\nu_{L}}.
\label{2.23f}
\end{align}
Consequently, given (\ref{2.20f}) the observed fractional change in the temperature in direction $\hat{n}$ at time $t=t_0$ is given to lowest perturbative order by 
\begin{align}
\frac{\Delta T(\hat{n},t_0)}{T_0}&=\frac{T(\hat{n},t_0)-T_0}{T_0}=
-1+\frac{\nu_0a(t_0)}{\nu_La(t_L)}+\frac{\delta T(r_L\hat{n},t_L)}{\bar{T}(t_L)}=\phi(r_L,t_L)-\phi(0,t_0)
\nonumber\\
&-\int_{t_L}^{t_0}dt\left(\frac{\partial N[r\hat{n},t]}{\partial t}\right)_{r=s(t)}
-a(t_L)\delta u^r(r_L\hat{n},t_L)+a(t_0)\delta u^r(0,t_0)+\frac{\delta T(r_L\hat{n},t_L)}{\bar{T}(t_L)}.
\label{2.24f}
\end{align}

To recast this expression we note that since (\ref{2.17f}) holds for any function of $r$ and $t$, we can set
\begin{align}
\frac{d}{dt}\left(a^2\frac{\partial^2 E}{\partial t^2}+a\frac{da}{dt}\frac{\partial E}{\partial t}+a\frac{\partial^2 E}{\partial t\partial r}\right)
&=\frac{\partial }{\partial t}\left(a^2\frac{\partial^2 E}{\partial t^2}+a\frac{da}{dt}\frac{\partial E}{\partial t}+a\frac{\partial^2 E}{\partial t\partial r}\right)
-\frac{1}{a}\frac{\partial}{\partial r}\left(a^2\frac{\partial^2 E}{\partial t^2}
+a\frac{da}{dt}\frac{\partial E}{\partial t}+a\frac{\partial^2 E}{\partial t\partial r}\right)
\nonumber\\
&=\frac{\partial}{\partial t}\left(a^2\frac{\partial^2 E}{\partial t^2}+a\frac{da}{dt}\frac{\partial E}{\partial t}\right)-\frac{\partial}{\partial r}\frac{\partial^2 E}{\partial t\partial r},
\nonumber \\
\frac{d}{dt}\left(a\frac{\partial E_r}{\partial t}-a\frac{\partial B}{\partial t}\right)&=\frac{\partial}{\partial t}\left(a\frac{\partial E_r}{\partial t}-a\frac{\partial B}{\partial t}\right)-\frac{1}{a}\frac{\partial}{\partial r}\left(a\frac{\partial E_r}{\partial t}-a\frac{\partial B}{\partial t}\right)
\nonumber\\
&=\frac{\partial}{\partial t}\left(a\frac{\partial E_r}{\partial t}-a\frac{\partial B}{\partial t}\right)-\frac{\partial}{\partial t}\left(\frac{\partial E_r}{\partial r}-\frac{\partial B}{\partial r}\right).
\label{2.25f}
\end{align}
Thus with $N[r\hat{n},t]$ being given in (\ref{2.6e}), we can rewrite (\ref{2.24f}) as
\begin{align}
\frac{\Delta T(\hat{n},t_0)}{T_0}&=\left(\phi-a^2\frac{\partial^2 E}{\partial t^2}-a\frac{da}{dt}\frac{\partial E}{\partial t}-a\frac{\partial^2 E}{\partial t\partial r}+a\frac{\partial B}{\partial t}-a\frac{\partial E_r}{\partial t}\right)\bigg{|}_{(r_L,t_L)}
-a(t_L)\delta u^r(r_L\hat{n},t_L)+\frac{\delta T(r_L\hat{n},t_L)}{\bar{T}(t_L)}
\nonumber\\
&+\int_{t_L}^{t_0}dt\frac{\partial}{\partial t}\left( \phi+B_r+\psi  - E_{rr}
-a^2\frac{\partial^2 E}{\partial t^2}-a\frac{da}{dt}\frac{\partial E}{\partial t}
+a\frac{\partial B}{\partial t}-a\frac{\partial E_r}{\partial t}
\right)_{r=s(t)}
\nonumber\\
&-\left(\phi-a^2\frac{\partial^2 E}{\partial t^2}-a\frac{da}{dt}\frac{\partial E}{\partial t}-a\frac{\partial^2 E}{\partial t\partial r}+a\frac{\partial B}{\partial t}-a\frac{\partial E_r}{\partial t }\right)\bigg{|}_{(0,t_0)}+a(t_0)\delta u^r(0,t_0),
\label{2.26f}
\end{align}
i.e., as
\begin{align}
\left(\frac{\Delta T(\hat{n})}{T_0}\right)_{\rm early}&=\left(\phi-a^2\frac{\partial^2 E}{\partial t^2}-a\frac{da}{dt}\frac{\partial E}{\partial t}-a\frac{\partial^2 E}{\partial t\partial r}+a\frac{\partial B}{\partial t}-a\frac{\partial E_r}{\partial t}\right)\bigg{|}_{(r_L,t_L)}
-a(t_L)\delta u^r(r_L\hat{n},t_L)+\frac{\delta T(r_L\hat{n},t_L)}{\bar{T}(t_L)},
\nonumber\\
\left(\frac{\Delta T(\hat{n})}{T_0}\right)_{\rm ISW}&=\int_{t_L}^{t_0}dt\frac{\partial}{\partial t}\left( \phi+B_r+\psi  - E_{rr}
-a^2\frac{\partial^2 E}{\partial t^2}-a\frac{da}{dt}\frac{\partial E}{\partial t}
+a\frac{\partial B}{\partial t}-a\frac{\partial E_r}{\partial t}
\right)_{r=s(t)},
\nonumber\\
\left(\frac{\Delta T(\hat{n})}{T_0}\right)_{\rm late}&=-\left(\phi-a^2\frac{\partial^2 E}{\partial t^2}-a\frac{da}{dt}\frac{\partial E}{\partial t}-a\frac{\partial^2 E}{\partial t\partial r}+a\frac{\partial B}{\partial t}-a\frac{\partial E_r}{\partial t }\right)\bigg{|}_{(0,t_0)}+a(t_0)\delta u^r(0,t_0).
\label{2.27f}
\end{align}
As we show in (\ref{3.9f}) and (\ref{3.10f}) below, if we set $B=0$, $B_r=0$, $E_r=0$, we can set $a(t)\delta u^r=\delta u_r/a(t)=\partial_r V/a(t)$. Thus, on setting $B=0$, $B_r=0$, $E_r=0$, $V=\delta u$ we recover (\ref{1.13e}) - (\ref{1.15e}), just as required, with (\ref{2.27f}) representing its generalization to the full set of SVT components in a $k=0$ background. We shall now discuss the structure of (\ref{2.27f}) and then generalize it to $k\neq 0$.

\section{The Fluid and Photon Velocities}
\label{S3}

In developing (\ref{2.27f}) we identified two types of velocities, the radial photon velocity and the background perfect fluid velocity. Even though the radial photon is part of the photon perfect fluid and moves with the fluid, the two types of velocity are distinct.  Specifically, the radial photon moves on the light cone and its velocity four-vector is lightlike. However, the background perfect fluid with form $T^{\mu\nu}=(\rho+p)u^{\mu}u^{\nu}+pg^{\mu\nu}$ is at rest in the comoving frame, and thus its four-vector is of the timelike form $u^{\mu}=(1,0,0,0)$ that obeys $g_{\mu\nu}u^{\mu}u^{\nu}=-1$. To reconcile these two types of velocities we note that a perfect fluid is actually an incoherent averaging of photons moving in all allowed directions (i.e., a statistical  average using a density matrix that is proportional to the unit matrix and normalized to one). Thus for the illustrative case (see e.g. \cite{Mannheim1988,Mannheim2006}) of a flat spacetime collection of massless plane waves each with a lightlike four-vector momentum $k^{\mu}$ that obeys $g_{\mu\nu}k^{\mu}k^{\nu}=0$, the energy-momentum tensor is of the traceless form $T^{\mu\nu}=k^{\mu}k^{\nu}/k^0$. With this $T^{\mu\nu}$ a plane wave $(k,0,0,k)$ moving in the positive $z$ direction contributes  
\begin{eqnarray}
T^{\mu\nu}(+z)=
\begin{pmatrix}
 k &  0 & 0&k \\ 
0& 0&0&0\\
0&0&0&0\\
k&0&0&k
\end{pmatrix}
\label{3.1e}
\end{eqnarray}
to $T^{\mu\nu}$. Incoherently adding a plane wave propagating in the negative $z$ direction gives 
\begin{eqnarray}
T^{\mu\nu}(+z)+T^{\mu\nu}(-z)=
\begin{pmatrix}
2 k &  0 & 0&0 \\ 
0& 0&0&0\\
0&0&0&0\\
0&0&0&2k
\end{pmatrix}.
\label{3.2e}
\end{eqnarray}
Finally, incoherently adding plane waves propagating in the positive and negative $x$ and $y$ directions we obtain 
\begin{eqnarray}
&&T^{\mu\nu}(+x)+T^{\mu\nu}(-x)+T^{\mu\nu}(+y)+T^{\mu\nu}(-y)+T^{\mu\nu}(+z)+T^{\mu\nu}(-z)=
\begin{pmatrix}
6k &  0 & 0&0 \\ 
0& 2k&0&0\\
0&0&2k&0\\
0&0&0&2k
\end{pmatrix},
\label{3.3e}
\end{eqnarray}
to thus be of the traceless perfect fluid form $T^{\mu\nu}=(\rho+p)u^{\mu}u^{\nu}+pg^{\mu\nu}$, where $u^{\mu}=(1,0,0,0)$ is timelike and where $\rho=3p$. An incoherent averaging of lightlike vectors thus generates a timelike one even as every mode in the photon fluid is still moving on the lightcone \cite{footnoteH}. A similar analysis holds in flat space polar coordinates \cite{Mannheim1988}. This analysis also  holds in the standard expanding Robertson-Walker background cosmologies with any spatial three-curvature, where it yields a photon energy density that behaves as $1/a^4(t)$ as a function of  $a(t)$  \cite{Mannheim1988}, with a thermodynamical incoherent statistical averaging at temperature $T$ yielding  $\rho=aT^4$ as function of  $T$ \cite{Deng1987}, just as required of the standard adiabatic $a(t)\sim 1/T$ relation \cite{footnoteH2}.

For fluctuations around the expanding Robertson-Walker background the same fluctuation dynamics that produces the $\delta \rho/4\rho=\delta T/T$ contribution that appears in (\ref{2.27f}) also produces fluctuations in the background perfect fluid four-velocity $u^{\mu}=(1,0,0,0)$. The $\delta T/T$ temperature fluctuations are not due to fluctuations in the fluid four-velocity, they are due to fluctuations in the fluid energy density, as per (\ref{2.22f}). However, fluctuations in the fluid energy density are of the same order as fluctuations in the fluid four-velocity as they are both associated with the same perturbed photon energy-momentum tensor. Fluid fluctuations thus generate a radial  $\delta u^r$ contribution to the fluctuating radial photon modes, and thus both of the $\delta \rho/\rho$ and $\delta u^r$ fluctuations contribute to the total $\Delta T/T_0$ that is given in (\ref{2.27f}). As as we show in (\ref{4.25g}) and (\ref{4.26g}), all of these fluid fluctuations are linked by gauge invariance, and one cannot have one without the other, with both of the $\delta \rho$ and $\delta u^r$ fluctuations being needed in order to obtain total temperature fluctuations $\Delta T/T_0$ that are gauge invariant. Consequently, while we have included both of these effects in the early time $\Delta T/T_0$, given the presence of a $\delta u^r$ contribution in the late time $\Delta T/T_0$, the gauge behavior of the perturbed photon energy momentum tensor that we describe below will require us to include a late time  $\delta \rho/\rho=4\delta T/T$ contribution as well.

In regard to these $\delta u^r$ fluctuations, we note that prior to last scattering the photon fluid is in thermodynamic equilibrium with the charged baryons and leptons that are continually absorbing and emitting photons until the temperature of the expanding and cooling Universe drops to the point where photons no longer have enough energy to ionize atoms as those atoms form. Because the photon, lepton and baryon fluids are all in thermodynamic equilibrium with each other,  all of the fluids possess a common timelike fluid four-velocity $u^{\mu}$. Thus even though the massive charged particles travel at less than the speed of light while the photons do travel at the speed of light,  they both are in perfect fluids that travel with the same $u^{\mu}$. Moreover, this remains the case for the fluctuations as well, and both massless and massive perturbed fluids have a common $\delta u^{\mu}$ \cite{Weinberg2008}. The $\delta u^r$ that appears in (\ref{2.27f}) is thus both the $\delta u^r$ of the particles emitting the photons and the $\delta u^r$ of the photon fluid as well. A similar situation prevails at the detector. Thus even while the photons are perturbed by  massive particles, they stay on the lightcone.

In (\ref{2.27f}) three eras are identified, viz.  early time, ISW and late time. In  terms of an actual CMB observation we note that in the early era charged particles emit photons, in the late era charged particles in detectors absorb these photons, and in the intermediate ISW era photons travel freely (modulo occasional collisions with intervening matter). Now both the emitter at last scattering and the detector of the current era observer are being swept along by the cosmic expansion, and to derive (\ref{2.27f}) we needed to compare waves that leave at $t_L$ and $t_L+\delta t_L$ with waves that arrive at the detector at $t_0$ and $t_0+\delta t_0$. We thus need to consider fluctuations in the fluid velocity at both emission and detection. (There are also fluctuations in the fluid velocity at all times between last scattering and the current era, but we are only considering photons that reach us without any substantive  intermediate era scattering). As we show below, we need to incorporate $\delta \rho$ and $\delta u^r$ fluid fluctuations at both the emitter and the detector in order to secure gauge invariance for both the early and late contributions in (\ref{2.27f}). However, since the intermediate era ISW photon propagation involves no interaction with charged fluid particles, we will find in (\ref{5.7f}) that the ISW contribution is purely geometric, with its gauge invariance being maintained purely by a gauge invariant interplay between the metric fluctuation components alone.

To explicitly incorporate the contribution of $\delta u^r$  into $\Delta T(\hat{n},t_0)/T_0$, we note that with all of the SVT terms being defined as covariants in (\ref{1.2e}), we need to determine the covariant component of the contravariant $\delta u^r$. With  $u^{\mu}=(1,0,0,0)$, $u_{\mu}=(-1,0,0,0)$, and with $h_{\mu\nu}$ given in (\ref{1.2e}) we obtain
\begin{align}
\delta u^r&=(g^{r\mu}-h^{r\mu})(u_{\mu}+\delta u_{\mu})-u^r=g^{rr}\delta u_{r}-h^{rt}u_{t}=\frac{1}{a^2(t)}\delta u_r
-\frac{1}{a(t)}(B_r+\partial_r B).
\label{3.4f}
\end{align}
(The contravariant $h^{r\mu}$ components appear with a minus sign in (\ref{3.4f}) in order to maintain $(g^{\mu\nu}-h^{\mu\nu})(g_{\nu\sigma}+h_{\nu\sigma})=\delta^{\mu}_{\sigma}$ to first order in the fluctuation.)
We note that the $B_r+\partial_r B$ term would be absent if the vector sector is not included, but as we will see below it will be needed in order to establish the gauge invariance of the general  (\ref{2.27f}).

However before proceeding to study gauge invariance we need to write $\delta u_r$ in a more convenient form, namely as 
the derivative of a scalar. To do this we note that $\delta u_r$ is the $r$ component of a covariant spatial three-vector $\delta u_i$. Thus as with any spatial three-vector, we can decompose $\delta u_i$ into transverse and longitudinal components  as $\delta u_i=V_i+\tilde{\nabla}_iV$, where $\tilde{\gamma}^{ij}\tilde{\nabla}_i V_j=0$ (see e.g. \cite{Phelps2019} for an explicit construction). For $k=0$ polar coordinates we have $\tilde{\gamma}_{ij}dx^idx^j=dr^2+r^2d\theta^2+r^2\sin^2\theta d\phi^2$, and with $\tilde{\gamma}^{1/2}=r^2\sin\theta$ we have
\begin{align}
\tilde{\gamma}^{ij}\tilde{\nabla}_iV_j=\tilde{\gamma}^{-1/2}\partial_i(\tilde{\gamma}^{1/2}\tilde{\gamma}^{ij}V_j)=0.
\label{3.5f}
\end{align}
So for purely radial motion with $\delta u_{\theta}=0$, $\delta u_{\phi}=0$, i.e., $V_{\theta}=-\partial_{\theta}V$, $V_{\phi}=-\partial_{\phi}V$, we have
\begin{align}
\frac{1}{r^2}\partial_r (r^2V_r)-\frac{1}{r^2\sin\theta}\partial_{\theta}\sin\theta\partial_{\theta}V-\frac{1}{r^2\sin^2\theta}\partial_{\phi}\partial_{\phi}V=0.
\label{3.6f}
\end{align}
Then if there is no dependence on the angular variables we have
\begin{align}
\frac{1}{r^2}\partial_r (r^2V_r)=0,
\label{3.7f}
\end{align}
with solution 
\begin{align}
V_r=\frac{c}{r^2}=-c\frac{\partial}{\partial r}\frac{1}{r},
\label{3.8f}
\end{align} 
where $c$ is a constant. As we see,  $V_r$ can be written as the derivative of a (three) scalar. But $\partial_rV$ is also the derivative of a scalar, so $\delta u_r$ can be written entirely as the derivative of a scalar \cite{footnoteI}. On setting $c$ to zero or on absorbing the $-c/r$ term into $V$ we can thus set $\delta u_r=\partial_rV$. Thus in terms of covariants we have 
\begin{align}
\delta u^r=\frac{1}{a^2(t)}\partial_r V
-\frac{1}{a(t)}(B_r+\partial_r B),\quad \delta u_r=\partial_r V.
\label{3.9f}
\end{align}

While this analysis constrains the radial dependence of $\delta u_r$, it does not affect its time dependence as we could multiply the above $V$ by an arbitrary function of the time. To fix such a possible overall multiplying factor it is convenient to 
switch to conformal time and the metric given in (\ref{1.8e}). For the conformal time metric the background perfect fluid velocity is given by $u^{\mu}=(1/\Omega(\tau),0,0,0)$, $u_{\mu}=(-\Omega(\tau),0,0,0)$. Thus for the conformal time background perfect fluid $T_{\mu\nu}=(\rho+p)u_{\mu}u_{\nu}+pg_{\mu\nu}$ we have $T_{00}=\rho\Omega^2(\tau)$, $T_{ij}=p\Omega^2(\tau)\tilde{\gamma}_{ij}$, to thus possess an overall factor of $\Omega^2(\tau)$. To understand the presence of this factor of $\Omega^2(\tau)$, we note that when restricted to $k=0$ the background metric in (\ref{1.8e}) is conformal to flat. And with a photon perfect fluid consisting of an incoherent average of massless modes all of which propagate on the light cone, the photon fluid is conformal invariant.  Thus we can start with a flat spacetime perfect fluid (i.e., one with constant $\rho$ and $p$) and make a conformal transformation on it to one propagating in the $k=0$ version of the metric (\ref{1.8e}). Under the local conformal transformation $g_{\mu\nu}(x)\rightarrow \Omega^2(x)g_{\mu\nu}(x)$ on the metric, $T_{\mu\nu}$ transforms as $\Omega^{-2}(x)T_{\mu\nu}$ (see e.g. \cite{Mannheim2006}). However, with the resulting photon $\rho$ and $p$ both behaving as $1/\Omega^4(\tau)$ \cite{footnoteJ}, to achieve a net $\Omega^{-2}(\tau)$, $u_{\mu}u_{\nu}$ must transform as $\Omega^2(\tau)$. When we now perturb the fluid, since all of its modes move on the perturbed light cone, the fluid remains conformal invariant (if $ds^2=(g_{\mu\nu}(x)+h_{\mu\nu}(x))dx^{\mu}dx^{\nu}=0$, then under $g_{\mu\nu}(x)+h_{\mu\nu}(x)\rightarrow \Omega^2(x)(g_{\mu\nu}(x)+h_{\mu\nu}(x))$ $ds^2$ remains zero). Since $g_{\mu\nu}u^{\mu}u^{\nu}=-1$ it follows that $\delta u_{\tau}=-\Omega(\tau)\phi$ (see e.g. \cite{Phelps2019}), where $\phi$ is given in the fluctuating part of (\ref{1.8e}). Thus $V$ must be proportional to $\Omega(\tau)$. Thus in the following we fix $V$ as $V=\Omega(\tau)X$. Since the transformation from conformal time to comoving time is just the coordinate transformation $d\tau=dt/a(t)$, in comoving time we set $V=a(t)X$, so that according to (\ref{3.9f}) in comoving time and conformal time we have 
\begin{align}
\delta u^r({\rm comoving})=\frac{1}{a(t)}\partial_r X-\frac{1}{a(t)}(B_r+\partial_r B),\quad \delta u_r({\rm comoving})=a(t)\partial_r X,
\nonumber\\
\delta u^r({\rm conformal})=\frac{1}{\Omega(\tau)}\partial_r X-\frac{1}{\Omega(\tau)}(B_r+\partial_r B),\quad \delta u_r({\rm conformal})=\Omega(\tau)\partial_r X.
\label{3.10f}
\end{align}
\section{Gauge Invariance and the Energy-Momentum Tensor}
\label{S4}

To explore the implications of gauge invariance for the energy-momentum tensor it is more convenient to first work in the conformal time fluctuation metric given in (\ref{1.8e}), a metric which applies for arbitrary $k$. For the background the energy-momentum tensor is of the perfect fluid form
\begin{align}
T_{\mu\nu}=(\rho+p)u_{\mu}u_{\nu}+pg_{\mu\nu},
\label{4.1g}
\end {align}
with $\rho(\tau)$ and $p(\tau)$ being functions of $\tau$ alone, functions that will be fixed once we introduce an equation of state that relates $\rho(\tau)$ and $p(\tau)$. However, for the moment we leave $\rho(\tau)$ and $p(\tau) $ to be general functions of $\tau$, with the gauge invariance analysis that we make in this section holding for both radiation and massive matter sources and not requiring the use of any equation of state that would relate $\rho(\tau)$ and $p(\tau)$. Under the fluctuation the change in $T_{\mu\nu}$  takes the form
\begin{align}
\delta T_{\mu\nu}=(\rho+p)[\delta u_{\mu}u_{\nu}+u_{\mu}\delta u_{\nu}]+ph_{\mu\nu}+(\delta \rho +\delta p)u_{\mu}u_{\nu}
+\delta p g_{\mu\nu},
\label{4.2g}
\end {align}
where $\delta u_{\tau}=-\Omega(\tau)\phi$, and where we leave $\delta u_i$ general for the moment as we are not yet restricting to radial motion alone, and to be as general as possible we instead first consider fluctuations associated with the arbitrary $\delta T_{\mu\nu}$. However, even in the general case we can still decompose $\delta u_i$ into transverse and longitudinal components as $\delta u_i=\Omega(\tau)(X_i+\tilde{\nabla}_iX)$ where $\tilde{\nabla}_iX^i=0$.

We now make a gauge transformation of the form $h_{\mu\nu}\rightarrow h_{\mu\nu}-\nabla _{\mu}\epsilon_{\nu}-\nabla_{\nu}\epsilon_{\mu}$ on $\delta T_{\mu\nu}$ so that $\delta T_{\mu\nu}\rightarrow \delta T_{\mu\nu}+\Delta \delta T_{\mu\nu}=\delta\bar{T}_{\mu\nu}$, where 
\begin{align}
\Delta \delta T_{\mu\nu}=\delta\bar{T}_{\mu\nu}-\delta T_{\mu\nu}=-T_{\lambda \mu}\partial_{\nu}\epsilon^{\lambda}-T_{\lambda \nu}\partial_{\mu}\epsilon^{\lambda}
-\epsilon^{\lambda}\partial_{\lambda}T_{\mu\nu}.
\label{4.3g}
\end{align}
With $u_{\tau}=-\Omega(\tau)$, on setting $\epsilon_{\tau}=-\Omega^2 T$, $\epsilon_i=\Omega^2(L_i+\tilde{\nabla}_iL)$ where $\tilde{\nabla}_iL^i=0$, and on recalling that the background $\rho$ and $p$ only depend on $\tau$, for the $(\tau\tau)$ component we obtain 
\begin{align}
\Delta \delta T_{\tau\tau}&=-2T_{\tau\tau}\partial_{\tau}\epsilon^{\tau}
-\epsilon^{\lambda}\partial_{\lambda}T_{\tau\tau}=-2\Omega^2\rho \dot{T}-2\rho \Omega\dot{\Omega}T-\Omega^2 T\dot{\rho},
\nonumber\\
\delta T_{\tau\tau}&=2\Omega^2\rho \phi+\Omega^2\delta \rho,\quad \delta\bar{T}_{\tau\tau}=2\Omega^2\rho \bar{\phi}+\Omega^2\delta\bar{\rho},
\nonumber\\
2\rho \bar{\phi}+\delta\bar{\rho}&=2\rho \phi+\delta \rho-2\rho \dot{T}-2\rho \Omega^{-1}\dot{\Omega}T-T\dot{\rho}.
\label{4.4g}
\end{align}
(Here the dot denotes the derivative with respect to $\tau$.) Now in \cite{Amarasinghe2018} we applied this same gauge transformation to the various SVT components, and in the $k=0$ case found that
\begin{align}
\bar{\phi}&=\phi-\dot{T}-\Omega^{-1}\dot{\Omega} T,\quad \bar{B}=B+T-\dot{L},\quad \bar{\psi}=\psi+\Omega^{-1}\dot{\Omega} T,
\nonumber\\
\bar{E}&=E-L,\quad \bar{B}_i=B_i-\dot{L}_i,\quad \bar{E}_i=E_i-L_i,\quad \bar{E}_{ij}=E_{ij}.
\label{4.5g}
\end{align}
In the Appendix we show that even with a non-zero background $k$ these relations continue to hold as is. Thus on using the relation for $\bar{\phi}$, from (\ref{4.4g}) we obtain 
\begin{align}
\delta\bar{\rho}=\delta \rho -T\dot{\rho}.
\label{4.6g}
\end{align}
With the covariant conservation of the background $T^{\mu\nu}$ leading to
\begin{align}
\dot{\rho}+3(p+\rho)\Omega^{-1}\dot{\Omega}=0,
\label{4.7g}
\end{align}
we can rewrite (\ref{4.6g}) as 
\begin{align}
\delta\bar{\rho}=\delta \rho +3T(\rho+p)\Omega^{-1}\dot{\Omega}.
\label{4.8g}
\end{align}
Thus on using the relation for $\bar{\psi}$ given in (\ref{4.5g}) we recognize
\begin{align}
\delta\hat{\rho}=\delta \bar{\rho} -3(\rho+p)\bar{\psi}=\delta \rho -3(\rho+p)\psi
\label{4.9g}
\end{align}
as being gauge invariant.

For  the $(i,\tau)$ sector we have 
\begin{align}
\Delta \delta T_{i\tau}&=-\rho\Omega^2\tilde{\nabla}_iT-p\Omega^2(\dot{L}_i+\tilde{\nabla}_i\dot{L}),
\nonumber\\
\delta T_{i\tau}&=-(\rho+p)\Omega^2(X_i+\tilde{\nabla}_iX)+p\Omega^2(B_i+\tilde{\nabla}_iB), 
\nonumber\\ 
\delta \bar{T}_{i\tau}&=-(\rho+p)\Omega^2(\bar{X}_i+\tilde{\nabla}_i\bar{X})+p\Omega^2(\bar{B}_i+\tilde{\nabla}_i\bar{B}).
\label{4.10g}
\end{align}
Thus we can set
\begin{align}
-(\rho+p)(\bar{X}_i+\tilde{\nabla}_i\bar{X})+p(\bar{B}_i+\tilde{\nabla}_i\bar{B})
=-(\rho+p)(X_i+\tilde{\nabla}_iX)+p(B_i+\tilde{\nabla}_iB)-\rho\tilde{\nabla}_iT-p(\dot{L}_i+\tilde{\nabla}_i\dot{L}).
\label{4.11g}
\end{align}
On applying $\tilde{\nabla}^i$ we thus obtain
\begin{align}
\tilde{\nabla}_i[-(\rho+p)(\bar{X}-X)+p(\bar{B}-B)+\rho T+p\dot{L}]=0,
\label{4.12g}
\end{align}
and ignoring a constant of integration set 
\begin{align}
-(\rho+p)(\bar{X}-X)+p(\bar{B}-B)+\rho T+p\dot{L}=0
\label{4.13g}
\end{align}
for the longitudinal sector. Inserting (\ref{4.12g}) into (\ref{4.11g}), for the transverse sector we obtain
\begin{align}
-(\rho+p)(\bar{X}_i-X_i)+p(\bar{B}_i-B_i)
+p\dot{L}_i=0.
\label{4.14g}
\end{align}
On using the relation for $\bar{B}$ given in (\ref{4.5g}), from (\ref{4.13g}) we find that $X$ transforms as
\begin{align}
\bar{X}=X+T
\label{4.15g}
\end{align}
under a gauge transformation. Similarly, on using the relation for $\bar{B}_i$ we find that 
\begin{align}
\bar{X}_i=X_i
\label{4.16g}
\end{align}
is gauge invariant. Finally, noting how $E$, $B$, $\psi$ and $X$ transform under a gauge transformation we find that
\begin{align}
\beta=\bar{B}-\dot{\bar{E}}-\bar{X}=B-\dot{E}-X,\quad \hat{X}=\bar{X}-\frac{\Omega}{\dot{\Omega}}\bar{\psi}=X-\frac{\Omega}{\dot{\Omega}}\psi=\gamma -\beta
\label{4.17g}
\end{align}
are gauge invariant, where $\gamma$ is given in (\ref{1.9e}). The $\hat{X}=\gamma-\beta$  relation is not an independent relation  but we list it here since as we show in the Appendix, $\hat{X}$ is the quantity that appears in the fluctuation Einstein equations \cite{footnoteK}. 

On recalling that the background $p$ only depends on $\tau$ and not on the spatial coordinates (so that $\tilde{\nabla}_ip=0$), for the $(i,j)$ sector we have 
\begin{align}
\Delta \delta T_{ij}&=-\Omega^2p\tilde{\gamma}_{ik}[f_m\partial_j\tilde{\gamma}^{km}+\tilde{\gamma}^{km}\partial_jf_m)
-\Omega^2p\tilde{\gamma}_{jk}[f_m\partial_i\tilde{\gamma}^{km}+\tilde{\gamma}^{km}\partial_if_m)
\nonumber\\
&-\tilde{\gamma}_{ij}T(\Omega^2\dot{p}+2\Omega\dot{\Omega}p)-\tilde{\gamma}^{km}\Omega^2p f_m\partial_k\tilde{\gamma}_{ij},
\label{4.18g}
\end{align}
where $f_m=L_m+\tilde{\nabla}_mL$. On taking the trace we obtain
\begin{align}
\tilde{\gamma}^{ij}\Delta \delta T_{ij}&=-2\Omega^2pf_m\partial_j\tilde{\gamma}^{jm}
-2\Omega^2p\tilde{\gamma}^{jm}\partial_jf_m
-3T(\Omega^2\dot{p}+2\Omega\dot{\Omega}p)-\tilde{\gamma}^{km}\Omega^2p f_m\tilde{\gamma}^{ij}\partial_k\tilde{\gamma}_{ij}.
\label{4.19g}
\end{align}
Noting that 
\begin{align}
\tilde{\gamma}^{jm}\tilde{\gamma}^{kn}\partial_j\tilde{\gamma}_{nm}=-\tilde{\gamma}^{jm}\tilde{\gamma}_{nm}\partial_j\tilde{\gamma}^{kn}=-\partial_j\tilde{\gamma}^{kj},
\label{4.20g}
\end{align}
and that
\begin{align}
\tilde{\gamma}^{jm}\tilde{\nabla}_jf_m&=\tilde{\gamma}^{jm}\partial_jf_m-\tilde{\gamma}^{jm}\tilde{\Gamma}^k_{jm}f_k=
\tilde{\gamma}^{jm}\partial_jf_m-\frac{1}{2}\tilde{\gamma}^{jm}\tilde{\gamma}^{kn}[\partial_j\tilde{\gamma}_{nm}+\partial_m\tilde{\gamma}_{nj}-\partial_n\tilde{\gamma}_{jm}]f_k
\nonumber\\
&=\tilde{\gamma}^{jm}\partial_jf_m+f_k\partial_j\tilde{\gamma}^{kj}+\frac{1}{2}f_k\tilde{\gamma}^{jm}\tilde{\gamma}^{kn}\partial_n\tilde{\gamma}_{jm},
\label{4.21g}
\end{align}
on recalling that $\tilde{\gamma}^{jm}\tilde{\nabla}_jL_m=0$ we obtain 
\begin{align}
\tilde{\gamma}^{ij}\Delta \delta T_{ij}&=
-2\Omega^2p\tilde{\gamma}^{jm}\tilde{\nabla}_jf_m
-3T(\Omega^2\dot{p}+2\Omega\dot{\Omega}p)=-2\Omega^2p\tilde{\nabla}^2L-3T(\Omega^2\dot{p}+2\Omega\dot{\Omega}p).
\label{4.22g}
\end{align}
Then, with $\tilde{\gamma}^{ij}\delta T_{ij}=p\Omega^2(-6\psi+2\tilde{\nabla}^2E)+3\delta p \Omega^2$ we obtain
\begin{align}
p(-6\bar{\psi}+2\tilde{\nabla}^2\bar{E})+3\delta\bar{p}=p(-6\psi+2\tilde{\nabla}^2E)+3\delta p 
-2p\tilde{\nabla}^2L-3T(\dot{p}+2\Omega^{-1}\dot{\Omega}p).
\label{4.23g}
\end{align}
Substituting the expressions for $\bar{\psi}$ and $\bar{E}$ given in (\ref{4.5g}) we find that 
\begin{align}
\delta \bar{p}=\delta p-T\dot{p},\quad \delta\hat{p}= \delta \bar{p}+\frac{\Omega}{\dot{\Omega}}\dot{p}\bar{\psi}=\delta p+\frac{\Omega}{\dot{\Omega}}\dot{p}\psi,
\label{4.24g}
\end{align}
with $\delta \hat{p}$ thus being gauge invariant. We should add that while we have only looked at the trace $\tilde{\gamma}^{ij}\Delta \delta T_{ij}$, there is of course more information contained in $\Delta \delta T_{ij}$ itself. However, as we show in the Appendix, this additional information is already contained in $\Delta\delta g_{ij}$. We also note that just as with the gauge invariant SVT combinations,  in deriving the gauge invariant energy-momentum tensor combinations we did not use any dynamical gravitational equation of motion. These relations are thus purely kinematic and hold in any pure metric-based theory of gravity.  

To summarize, in conformal time the full set of gauge invariants are 
\begin{align}
&\alpha=\phi + \psi + \frac{\partial B}{\partial\tau} - \frac{\partial^2E}{\partial\tau^2},\quad  \gamma= - \left(\frac{d \Omega}{d \tau}\right)^{-1}\Omega \psi + B - \frac{\partial E}{\partial\tau},\quad  B_i-\frac{\partial E_i}{\partial \tau},  \quad E_{ij},
\nonumber\\
&\delta\hat{\rho}=\delta \rho -3(\rho+p)\psi,\quad X_i,\quad \hat{X}=X-\left(\frac{d \Omega}{d \tau}\right)^{-1}\Omega\psi,\quad \beta=B-\frac{\partial E}{\partial \tau}-X,
\quad
\delta \hat{p}=\delta p+\left(\frac{d\Omega}{d\tau }\right)^{-1}\Omega\frac{dp}{d\tau}\psi,
\label{4.25g}
\end{align}
and in comoving time they are
\begin{align}
&\alpha=\phi + \psi + a\frac{\partial B}{\partial t} - a^2\frac{ \partial^2E}{\partial t^2}-a\frac{da}{dt} \frac{\partial E}{\partial t},\quad  \gamma= - \left(\frac{da }{dt}\right)^{-1}\psi + B - a\frac{\partial E}{\partial t},\quad  B_i-a\frac{\partial E_i}{ \partial t},  \quad E_{ij},
\nonumber\\
&\delta\hat{\rho}=\delta \rho -3(\rho+p)\psi,\quad X_i,\quad \hat{X}=X-\left(\frac{d a}{d t}\right)^{-1}\psi,\quad \beta=B-a\frac{\partial E}{\partial t}-X,
\quad
\delta \hat{p}=\delta p+a\left(\frac{da}{dt}\right)^{-1}\frac{dp}{dt}\psi.
\label{4.26g}
\end{align}
We note that though these relations were derived in a space with non-vanishing spatial three-curvature $k$, these relations have no dependence on $k$. They thus hold in an unmodified form for any value of $k$ \cite{footnoteL}. To confirm the validity of these relations, in the Appendix we evaluate the gauge invariant Einstein gravity fluctuation quantity $\Delta_{\mu\nu}=\delta G_{\mu\nu}+8\pi G\delta T_{\mu\nu}$ in the arbitrary $k$ case,  and show that it is composed solely of these specific $k$-independent gauge invariant combinations. Armed with these gauge invariant combinations we now show that the light cone temperature fluctuations are gauge invariant.

\section{Gauge Invariance of the Light Cone $k=0$ Temperature Fluctuation Relations in Comoving Time}
\label{S5}

Since the above gauge invariant relations were obtained for general $\delta u_i$ and general $k$, they also hold when $i=r$ and $k=0$. We can thus use them for the $k=0$ temperature fluctuations we derived above. Comparison with (\ref{1.6e}) shows that we can immediately write the comoving time ISW term in (\ref{2.27f}) in a manifestly gauge invariant form 
\begin{align}
\left(\frac{\Delta T(\hat{n})}{T_0}\right)_{\rm ISW}=\int_{t_L}^{t_0}dt\frac{\partial}{\partial t}\left( \alpha+B_r  -a\frac{\partial E_r}{\partial t}- E_{rr}\right)_{r=s(t)},
\label{5.1e}
\end{align}
and we note that the ISW contribution is a purely geometric contribution that only depends on the SVT combinations and not on any of the combinations associated with the energy-momentum tensor. Also we note that the SVT $\gamma$ combination given in (\ref{1.6e}) is not included in (\ref{5.1e}), a point we return to below. For the comoving time early and late contributions in (\ref{2.27f}), on recalling (\ref{3.10f}) we generically set
\begin{align}
&\phi-a^2\frac{\partial^2 E}{\partial t^2}-a\frac{da}{dt}\frac{\partial E}{\partial t}-a\frac{\partial^2 E}{\partial t\partial r}+a\frac{\partial B}{\partial t}-a\frac{\partial E_r}{\partial t}-a\delta u^r+\frac{\delta T}{T}
\nonumber\\
&=\alpha+B_r-a\frac{\partial E_r}{\partial t}+\frac{\partial}{\partial r}\left[-X+B-a\frac{\partial E}{\partial t}\right] -\psi+\frac{\delta T}{T}
\nonumber\\
&=\alpha+B_r-a\frac{\partial E_r}{\partial t}+\frac{\partial \beta}{\partial r}-\psi+\frac{\delta \rho}{4\rho},
\label{5.2e}
\end{align}
where in the last line we have used (\ref{4.26g}) and (\ref{2.22f}). Now while the $\delta \hat{\rho}=\delta \rho -3(\rho+p)\psi$ relation given in (\ref{4.26g}) is general, for radiation it reduces to $\delta \hat{\rho}=\delta \rho -4\rho\psi$. Thus for the early contribution we obtain
\begin{align}
\left(\frac{\Delta T(\hat{n})}{T_0}\right)_{\rm early}&=\left[\alpha+B_r-a\frac{\partial E_r}{\partial t}+\frac{\partial \beta}{\partial r}+\frac{\delta \hat{\rho}}{4\rho} \right]\bigg{|}_{(r_L,t_L)},
\label{5.3f}
\end{align}
and thus establish that it is gauge invariant. 

However, as it stands the late value as given in (\ref{2.27f}) is not gauge invariant as it is given by
\begin{align}
\left(\frac{\Delta T(\hat{n})}{T_0}\right)_{\rm late}&=-\left[\alpha+B_r-a\frac{\partial E_r}{\partial t}+\frac{\partial \beta}{\partial r}-\psi \right]\bigg{|}_{(0,t_0)}.
\label{5.4f}
\end{align}
To rectify this we must, as discussed  in Sec. \ref{S3},  also consider a temperature fluctuation at the observer (viz. a change in the photon energy density at the observer) by replacing the late part of (\ref{2.27f}) by 
\begin{align}
\left(\frac{\Delta T(\hat{n})}{T_0}\right)_{\rm late}&=-\left(\phi-a^2\frac{\partial^2 E}{\partial t^2}-a\frac{da}{dt}\frac{\partial E}{\partial t}-a\frac{\partial^2 E}{\partial t\partial r}+a\frac{\partial B}{\partial t}-a\frac{\partial E_r}{\partial t }\right)\bigg{|}_{(0,t_0)}+a(t_0)\delta u^r(0,t_0)-\frac{\delta T(0,t_0)}{T_0}.
\label{5.5f}
\end{align}
With this modification we can now set 
\begin{align}
\left(\frac{\Delta T(\hat{n})}{T_0}\right)_{\rm late}&=-\left[\alpha+B_r-a\frac{\partial E_r}{\partial t}+\frac{\partial \beta}{\partial r} +\frac{\delta \hat{\rho}}{4\rho} \right]\bigg{|}_{(0,t_0)},
\label{5.6f}
\end{align}
to now be manifestly gauge  invariant. Thus to summarize, in comoving time we can write the full $\Delta T(\hat{n},t_0)/T_0$ that the observer at $r=0$, $t=t_0$ sees is 
\begin{align}
\frac{\Delta T(\hat{n},t_0)}{T_0}&=\left[\alpha+B_r-a\frac{\partial E_r}{\partial t}+\frac{\partial \beta}{\partial r} +\frac{\delta \hat{\rho}}{4\rho} \right]\bigg{|}_{(r_L,t_L)}
+\int_{t_L}^{t_0}dt\frac{\partial}{\partial t}\left( \alpha+B_r  -a\frac{\partial E_r}{\partial t}- E_{rr}\right)_{r=s(t)}
\nonumber\\
&-\left[\alpha+B_r-a\frac{\partial E_r}{\partial t}+\frac{\partial \beta}{\partial r}+\frac{\delta \hat{\rho}}{4\rho} \right]\bigg{|}_{(0,t_0)},
\label{5.7f}
\end{align}
a now fully gauge invariant form \cite{footnoteM}.

An interesting feature of these relations is that neither the early nor the late expressions would be gauge invariant if we had not included the $\delta u^r$ term. Specifically, as given in (\ref{3.10f}) $a\delta u^r=\partial_r(X-B)-B_r$ is not on its own gauge invariant. And nor for that matter is $-a\delta u^r+\delta T/T$. Thus if we were to leave $-a\delta u^r$ out, what would remain would not be gauge invariant. A quick way to see this is to note that we need a $B_r$ factor to augment each $-a\partial_tE_r$ term in (\ref{5.7f}) and it can only come from $\delta u^r$. However in contrast, we note that just as we had expressly noted in Sec. \ref{S3} and just as we now explicitly see in (\ref{5.7f}), the ISW term is not dependent on $\delta u^r$ at all, with its gauge invariance being maintained by the metric fluctuations alone. The ISW fluctuation term is thus purely geometric.

\section{Gauge Invariance of the Light Cone $k=0$ Temperature Fluctuation Relations in Conformal Time}
\label{S6}

Since the transformation from comoving time to conformal time is just a coordinate transformation $d\tau=dt/a(t)$ and since temperature is a general coordinate scalar we can directly transcribe (\ref{5.7f}) to 
\begin{align}
\frac{\Delta T(\hat{n},\tau_0)}{T_0}&=\left[\alpha+B_r-\frac{\partial E_r}{\partial \tau}+\frac{\partial \beta}{\partial r} +\frac{\delta \hat{\rho}}{4\rho} \right]\bigg{|}_{(r_L,\tau_L)}
+\int_{\tau_L}^{\tau_0}d\tau\frac{\partial}{\partial \tau}\left( \alpha+B_r  -\frac{\partial E_r}{\partial \tau}- E_{rr}\right)_{r=s(\tau)}
\nonumber\\
&-\left[\alpha+B_r-\frac{\partial E_r}{\partial \tau}+\frac{\partial \beta}{\partial r}+\frac{\delta \hat{\rho}}{4\rho} \right]\bigg{|}_{(0,\tau_0)},
\label{6.1f}
\end{align}
where now 
\begin{align}
&\alpha=\phi + \psi + \frac{\partial B}{\partial\tau} - \frac{\partial^2E}{\partial\tau^2},\quad   B_r-\frac{\partial E_r}{\partial \tau},  \quad E_{rr},
\nonumber\\
&\delta\hat{\rho}=\delta \rho -3(\rho+p)\psi,\quad \beta=B-\frac{\partial E}{\partial \tau}-X
\label{6.2f}
\end{align}
are the relevant gauge invariants. Missing from the temperature fluctuations are the other gauge invariants:
\begin{align}
& \gamma= - \left(\frac{d \Omega}{d \tau}\right)^{-1}\Omega \psi + B - \frac{\partial E}{\partial\tau},\quad  X_i,\quad \hat{X}=X-\left(\frac{d \Omega}{d \tau}\right)^{-1}\Omega\psi,\quad
\delta \hat{p}=\delta p+\Omega\left(\frac{d\Omega}{d\tau }\right)^{-1}\frac{dp}{d\tau}\psi.
\label{6.3f}
\end{align}
Now the three-vector $X_i$ is absent simply because we only considered radial modes in which $\delta u_r$ is the derivative of a scalar. However, the other absent combinations all involve the conformal factor while none of the combinations that are present in (\ref{6.1f}) do. This is to be expected since the light cone is conformal invariant, i.e., under $g_{\mu\nu}(x)\rightarrow e^{2\alpha(x)}g_{\mu\nu}(x)$ the line element $ds^2=-g_{\mu\nu}dx^{\mu}dx^{\nu}$ transforms into $e^{2\alpha(x)}ds^2$, and is thus left invariant if $ds^2=0$. Since the light cone is conformal invariant, multiplying the metric by a conformal factor cannot change light cone fluctuations. Hence the temperature fluctuations must be independent of $\Omega(\tau)$, just as we find them to be, with only $\Omega(\tau)$-independent combinations appearing in the temperature fluctuations exhibited in (\ref{6.1f}). We provide some further insight into these conformal issues in the Appendix.

\section{Gauge Invariance of the Light Cone $k\neq 0$ Temperature Fluctuation Relations in Conformal or Comoving Time}
\label{S7}

To discuss the $k \neq 0$ case we recall that even though the gauge invariant combinations that are given in (\ref{4.25g}) and (\ref{4.26g}) were derived in a $k\neq 0$ background they actually have no explicit dependence $k$. We thus anticipate that this will also be true of the comoving (\ref{5.7f}) and conformal (\ref{6.1f}) temperature fluctuation relations themselves. To establish that the temperature fluctuation relations do continue to hold without modification, we rewrite the conformal time metric (\ref{1.8e}) using $r=\sinh \chi$ when $k<0$ and using $r=\sin\chi$ when $k>0$. For the $k<0$ case first this leads to 
\begin{align}
ds^2&=\Omega^2(\tau)\left[d\tau^2-d\chi^2-\sinh^2\chi d\theta^2-\sinh^2\chi\sin^2\theta d\phi^2\right]
\nonumber\\
&+\Omega^2(\tau)\left[2\phi d\tau^2 -2(\tilde{\nabla}_i B +B_i)d\tau dx^i - [-2\psi\tilde{\gamma}_{ij} +2\tilde{\nabla}_i\tilde{\nabla}_j E + \tilde{\nabla}_i E_j + \tilde{\nabla}_j E_i + 2E_{ij}]dx^i dx^j\right],
\label{7.1h}
\end{align}
where $(1,2,3)$ denote $(\chi,\theta,\phi)$. Radial photon modes are now those that satisfy
\begin{align}
ds^2&=\Omega^2(\tau)\left[d\tau^2-d\chi^2\right]+\Omega^2(\tau)\left[2\phi d\tau^2 -2(\partial_{\chi}B +B_{\chi})d\tau d\chi - [-2\psi +2\partial_{\chi}\partial_{\chi} E + 2\partial_{\chi}E_{\chi}+ 2E_{\chi\chi}]d\chi^2\right]=0.
\label{7.2h}
\end{align}
We note that (\ref{7.2h}) is completely analogous to the conformal time variant of (\ref{2.2e}) with $\chi$ having replaced $r$, and with the background modes obeying $d\chi/d\tau=-1$ in analog to the previous discussion where $dr/dt=-1/a(t)$, $dr/d\tau=-1$. Now we recall that (\ref{2.2e}) was actually developed for $k=0$. However, since there is no explicit $k$ dependence in (\ref{7.2h}) even though (\ref{7.2h}) applies in the non-zero $k$ case, it follows that the temperature fluctuations associated with (\ref{7.2h}) should have no explicit dependence on $k$ either, just as we now show. 

However, in order to specifically extend the analysis to non-zero $k$ we need to generalize the analysis of $\delta u^r$ given in Sec. \ref{S3}. Thus we set
\begin{align}
\delta u^{\chi}&=(g^{\chi\mu}-h^{\chi\mu})(u_{\mu}+\delta u_{\mu})-u^{\chi}=g^{\chi\chi}\delta u_{\chi}-h^{\chi \tau}u_{\tau}=\frac{1}{\Omega^2(\tau)}\delta u_{\chi}
-\frac{1}{\Omega(\tau)}(B_{\chi}+\partial_{\chi} B).
\label{7.3h}
\end{align}
As before we set $\delta u_i=V_i+\tilde{\nabla}_iV$, where $\tilde{\gamma}^{ij}\tilde{\nabla}_i V_j=0$ and where now $\tilde{\gamma}^{1/2}=\sinh^2\chi\sin\theta$. Thus for pure radial motion with $\delta u_{\chi}=V_{\chi}+\partial_{\chi}V$, we obtain
\begin{align}
\tilde{\gamma}^{-1/2}\partial_i(\tilde{\gamma}^{1/2}\tilde{\gamma}^{ij}V_j)=\frac{1}{\sinh^2\chi}\partial_{\chi} (\sinh^2\chi V_{\chi})=0,
\label{7.4h}
\end{align}
with solution 
\begin{align}
V_{\chi}=\frac{c}{\sinh^2\chi}=-c\frac{\partial}{\partial \chi}\frac{\cosh\chi}{\sinh \chi},
\label{7.5h}
\end{align} 
where $c$ is a constant. As we see,  $V_{\chi}$ can be written as the derivative of a (three) scalar. But $\partial_{\chi}V$ is also the derivative of a scalar, so $\delta u_{\chi}$ can be written entirely as the derivative of a scalar. On setting $c$ to zero or on absorbing the $-c\cosh\chi/\sinh\chi$ term into $V$ we can thus set $\delta u_{\chi}=\partial_{\chi}V$. And on setting $V=\Omega(\tau)X$ we have 
\begin{align}
\delta u^{\chi}=\frac{1}{\Omega(\tau)}\partial_{\chi} X
-\frac{1}{\Omega(\tau)}(B_{\chi}+\partial_{\chi}B),\quad \delta u_{\chi}=\Omega(\tau)\partial_{\chi}X,
\label{7.6h}
\end{align}
the complete analog of the conformal time (\ref{3.10f}). 

For  radial modes with $k>0$ everything is the same, with $\sinh\chi$ being replaced by $\sin\chi$ in (\ref{7.1h}), with (\ref{7.2h}) still holding, and with (\ref{7.4h}) and (\ref{7.5h}) being replaced by
\begin{align}
\frac{1}{\sin^2\chi}\partial_{\chi} (\sin^2\chi V_{\chi})=0,
\label{7.7h}
\end{align}
and
\begin{align}
V_{\chi}=\frac{c}{\sin^2\chi}=-c\frac{\partial}{\partial \chi}\frac{\cos\chi}{\sin \chi}.
\label{7.8h}
\end{align} 
Thus again $V_{\chi}$ is the derivative of a (three) scalar, and thus again we obtain (\ref{7.6h}).

Thus from this point on for both $k<0$ and $k>0$ the discussion completely parallels the $k=0$ discussion with $r$ replaced everywhere by $\chi$. And on setting $s(\tau)$ and $s(t)$ to be the background trajectories that start at $(\chi_L,\tau_L)$ or $(\chi_L,t_L)$, the conformal time (\ref{6.1f}) and comoving time (\ref{5.7f}) will hold as is even if the three-curvature $k$ is non-zero, being of the form:
\begin{align}
\frac{\Delta T(\hat{n},\tau_0)}{T_0}&=\left[\alpha+B_{\chi}-\frac{\partial E_{\chi}}{\partial \tau}+\frac{\partial \beta}{\partial \chi} +\frac{\delta \hat{\rho}}{4\rho} \right]\bigg{|}_{(\chi_L,\tau_L)}
+\int_{\tau_L}^{\tau_0}d\tau\frac{\partial}{\partial \tau}\left( \alpha+B_{\chi}  -\frac{\partial E_{\chi}}{\partial \tau}- E_{\chi\chi}\right)_{\chi=s(\tau)}
\nonumber\\
&-\left[\alpha+B_{\chi}-\frac{\partial E_{\chi}}{\partial \tau}+\frac{\partial \beta}{\partial \chi}+\frac{\delta \hat{\rho}}{4\rho} \right]\bigg{|}_{(0,\tau_0)},
\label{7.9h}
\end{align}
and
\begin{align}
\frac{\Delta T(\hat{n},t_0)}{T_0}&=\left[\alpha+B_{\chi}-a\frac{\partial E_{\chi}}{\partial t}+\frac{\partial \beta}{\partial \chi} +\frac{\delta \hat{\rho}}{4\rho} \right]\bigg{|}_{(\chi_L,t_L)}
+\int_{t_L}^{t_0}dt\frac{\partial}{\partial t}\left( \alpha+B_{\chi}  -a\frac{\partial E_{\chi}}{\partial t}- E_{\chi\chi}\right)_{\chi=s(t)}
\nonumber\\
&-\left[\alpha+B_{\chi}-a(t)\frac{\partial E_{\chi}}{\partial t}+\frac{\partial \beta}{\partial {\chi}}+\frac{\delta \hat{\rho}}{4\rho} \right]\bigg{|}_{(0,t_0)}.
\label{7.10h}
\end{align}
This then is our main result. 

To understand why there is no explicit $k$ dependence in (\ref{7.9h}) and (\ref{7.10h}), it is instructive to note that by a sequence of general coordinate transformations one can write the background metric 
\begin{eqnarray}
ds^2&=&a^2(\tau)\left[d\tau^2-\frac{dr^2}{1-kr^2}-r^2d\theta^2-r^2\sin^2\theta d\phi^2\right]
\label{7.11h} 
\end{eqnarray}
in a conformal to flat form. Specifically in the $k<0$ case first, it is convenient to set $k=-1/L^2$, and introduce ${\rm sinh} \chi=r/L$ and $p=\tau/L$, with the  metric given in (\ref{7.11h}) then taking the form
\begin{eqnarray}
ds^2=L^2a^2(p)\left[dp^2-d\chi^2 -{\rm sinh}^2\chi d\theta^2-{\rm sinh}^2\chi \sin^2\theta d\phi^2\right].
\label{7.12h}
\end{eqnarray}
Next we introduce  (see e.g. \cite{Amarasinghe2018})
\begin{eqnarray}
p^{\prime}+r^{\prime}=\tanh[(p+\chi)/2],\quad p^{\prime}-r^{\prime}=\tanh[(p-\chi)/2],\quad p^{\prime}=\frac{\sinh p}{\cosh p+\cosh \chi},\quad r^{\prime}=\frac{\sinh \chi}{\cosh p+\cosh \chi},
\label{7.13h}
\end{eqnarray}
so that
\begin{eqnarray}
dp^{\prime 2}-dr^{\prime 2}&=&\frac{1}{4}[dp^2-d\chi^2]{\rm sech}^2[(p+\chi)/2]{\rm sech}^2[(p-\chi)/2],
\nonumber\\
\frac{1}{4}(\cosh p+\cosh \chi)^2&=&{\rm \cosh}^2[(p+\chi)/2]{\rm \cosh}^2[(p-\chi)/2]=\frac{1}{[1-(p^{\prime}+r^{\prime})^2][1-(p^{\prime}-r^{\prime})^2]}.
\label{7.14h}
\end{eqnarray}
With these transformations the line element takes the conformal to flat form
\begin{eqnarray}
ds^2=\frac{4L^2a^2(p)}{[1-(p^{\prime}+r^{\prime})^2][1-(p^{\prime}-r^{\prime})^2]}\left[dp^{\prime 2}-dr^{\prime 2} -r^{\prime 2}d\theta^2-r^{\prime 2} \sin^2\theta d\phi^2\right].
\label{7.15h}
\end{eqnarray}
To bring the spatial sector  of (\ref{7.15h}) to Cartesian coordinates we set  $x^{\prime}=r^{\prime}\sin\theta\cos\phi$, $y^{\prime}=r^{\prime}\sin\theta\sin\phi$, $z^{\prime}=r^{\prime}\cos\theta$, $r^{\prime}=(x^{\prime 2}+ y^{\prime 2}+z^{\prime 2})^{1/2}$, and thus bring the line element to the form  
\begin{eqnarray}
ds^2=L^2a^2(p)(\cosh p+\cosh \chi)^2\left[dp^{\prime 2}-dx^{\prime 2} -dy^{\prime 2} -dz^{\prime 2}\right].
\label{7.16h}
\end{eqnarray}
With these transformations the $k<0$ line element is now in a conformal to flat form.

For the $k>0$ case the situation is analogous. Starting with conformal time (\ref{7.11h}) 
we set  $k=1/L^2$ and $\sin \chi=r/L$, with the conformal time metric  then taking the form
\begin{eqnarray}
ds^2=L^2a^2(p)\left[dp^2-d\chi^2 -\sin^2\chi d\theta^2-\sin^2\chi \sin^2\theta d\phi^2\right],
\label{7.17h}
\end{eqnarray}
where $p=\tau/L$. Following e.g.  \cite{Amarasinghe2018} we introduce
\begin{eqnarray}
p^{\prime}+r^{\prime}=\tan[(p+\chi)/2],\qquad p^{\prime}-r^{\prime}=\tan[(p-\chi)/2],\qquad p^{\prime}=\frac{\sin p}{\cos p+\cos \chi},\qquad r^{\prime}=\frac{\sin \chi}{\cos p+\cos \chi},
\label{7.18h}
\end{eqnarray}
so that
\begin{eqnarray}
dp^{\prime 2}-dr^{\prime 2}&=&\frac{1}{4}[dp^2-d\chi^2]\sec^2[(p+\chi)/2]\sec^2[(p-\chi)/2],
\nonumber\\
\frac{1}{4}(\cos p +\cos \chi)^2&=&\cos^2[(p+\chi)/2]\cos^2[(p-\chi)/2]=\frac{1}{[1+(p^{\prime}+r^{\prime})^2][1+(p^{\prime}-r^{\prime})^2]}.
\label{7.19h}
\end{eqnarray}
With these transformations the $k>0$ line element then takes the conformal to flat form
\begin{eqnarray}
ds^2&=&\frac{4L^2a^2(p)}{[1+(p^{\prime}+r^{\prime})^2][1+(p^{\prime}-r^{\prime})^2]}\left[dp^{\prime 2}-dr^{\prime 2} -r^{\prime 2}d\theta^2-r^{\prime 2} \sin^2\theta d\phi^2\right]
\nonumber\\
&=&L^2a^2(p)(\cos p+\cos \chi)^2\left[dp^{\prime 2}-dx^{\prime 2} -dy^{\prime 2} -dz^{\prime 2} \right].
\label{7.21h}
\end{eqnarray}

Since the entire dependence on the spatial three-curvature can be put into the conformal factors exhibited in (\ref{7.16h}) and (\ref{7.21h}), and since the light cone temperature fluctuations are independent of any overall conformal factor, it follows that the light cone temperature fluctuations must be independent of the three-curvature $k$, just as we have found. Consequently, the light cone temperature fluctuation relations given in (\ref{7.9h}) and (\ref{7.10h}) must have no explicit dependence on either the conformal factor or the spatial three-curvature, a quite remarkable simplification.

\begin{acknowledgments}
The authors acknowledge useful conversations with T. Liu and  D. Norman. 
\end{acknowledgments}

\appendix
\numberwithin{equation}{section}
\setcounter{equation}{0}

\section{Gauge Structure of the Metric Fluctuations}
\label{SA}
The transformation relations given in (\ref{4.5g}) were derived in \cite{Amarasinghe2018} for $k=0$. We now generalize them to non-zero $k$. The structure of fluctuations in the metric sector parallels the structure of fluctuations in the energy-momentum tensor sector as both are rank two tensors, with (\ref{4.3g}) thus being replaced by
\begin{align}
\Delta \delta g_{\mu\nu}=\delta\bar{g}_{\mu\nu}-\delta g_{\mu\nu}=-g_{\lambda \mu}\partial_{\nu}\epsilon^{\lambda}-g_{\lambda \nu}\partial_{\mu}\epsilon^{\lambda}
-\epsilon^{\lambda}\partial_{\lambda}g_{\mu\nu}.
\label{A.1}
\end{align}
Recalling that $g_{00}$ is negative, we can transcribe the relations we obtained for $\Delta \delta T_{\mu\nu}$ by setting $\rho=-1$, $p=1$. Thus for the conformal time $(\tau,\tau)$ component we obtain
\begin{align}
\Delta \delta g_{\tau\tau}&=2\Omega^2 \dot{T}+2 \Omega\dot{\Omega}T.
\label{A.2}
\end{align}
(Here the dot denotes the derivative with respect to $\tau$.) With $h_{\tau\tau}$ and $\bar{h}_{\tau\tau}$ being of the form
\begin{align}
h_{\tau\tau}&=-2\Omega^2 \phi,\quad \bar{h}_{\tau\tau}=-2\Omega^2 \bar{\phi},
\label{A.3}
\end{align}
we obtain
\begin{align}
\bar{\phi}=\phi-\dot{T} - \Omega^{-1}\dot{\Omega}T.
\label{A.4}
\end{align}

For  the $(i,\tau)$ sector we have 
\begin{align}
\Delta \delta g_{i\tau}&=\Omega^2\tilde{\nabla}_iT-\Omega^2(\dot{L}_i+\tilde{\nabla}_i\dot{L}),
\quad
h_{i\tau}=\Omega^2(B_i+\tilde{\nabla}_iB),\quad  \bar{h}_{i\tau}=\Omega^2(\bar{B}_i+\tilde{\nabla}_i\bar{B}).
\label{A.5}
\end{align}
Thus we obtain
\begin{align}
\bar{B}_i+\tilde{\nabla}_i\bar{B}=B_i+\tilde{\nabla}_iB+\tilde{\nabla}_iT-\dot{L}_i-\tilde{\nabla}_i\dot{L}.
\label{A.6}
\end{align}
On applying $\tilde{\nabla}^i$ we thus obtain
\begin{align}
\tilde{\nabla}^2(\bar{B}-B-T+\dot{L})=0.
\label{A.7}
\end{align}
Now in \cite{Phelps2019} and \cite{Mannheim2020} generic equations of the form $(\tilde{\nabla}^2+A_S)S=0$  were studied where $A_S$ is a numerical constant and $S$ is a generic scalar field,  and in the $k<0$ case (which we discuss first) it was shown that under the boundary conditions that $S$ vanish at $r=\infty$ and be well-behaved at $r=0$ the only allowed solution with $A_S=0$ is $S=0$. Thus under these boundary conditions we obtain
\begin{align}
\bar{B}=B+T-\dot{L}
\label{A.8}
\end{align}
for the longitudinal sector. Inserting (\ref{A.8}) into (\ref{A.6}), for the transverse sector we obtain
\begin{align}
\bar{B}_i=B_i-\dot{L}_i.
\label{A.9}
\end{align}

For the $(i,j)$ sector we have 
\begin{align}
\Delta \delta g_{ij}&=-\Omega^2\tilde{\gamma}_{ik}(f_m\partial_j\tilde{\gamma}^{km}+\tilde{\gamma}^{km}\partial_jf_m)
-\Omega^2\tilde{\gamma}_{jk}(f_m\partial_i\tilde{\gamma}^{km}+\tilde{\gamma}^{km}\partial_if_m)
-2\tilde{\gamma}_{ij}T\Omega\dot{\Omega}-\tilde{\gamma}^{km}\Omega^2 f_m\partial_k\tilde{\gamma}_{ij},
\label{A.10}
\end{align}
where $f_m=L_m+\tilde{\nabla}L$. We can rewrite this expression
\begin{align}
\Omega^{-2}\Delta \delta g_{ij}&=\tilde{\gamma}^{km}f_m\partial_j\tilde{\gamma}_{ik}-\partial_jf_i
+\tilde{\gamma}^{km}f_m\partial_i\tilde{\gamma}_{jk}-\partial_if_j
-2\tilde{\gamma}_{ij}T\Omega^{-1}\dot{\Omega}-\tilde{\gamma}^{km} f_m\partial_k\tilde{\gamma}_{ij}.
\label{A.11}
\end{align}
Noting that
\begin{align}
\tilde{\nabla}_if_j&=\partial_if_j-\frac{1}{2}\tilde{\gamma}^{km}(\partial_i\tilde{\gamma}_{mj}+\partial_j\tilde{\gamma}_{mi}
-\partial_m\tilde{\gamma}_{ij})f_{k},
\nonumber\\
\tilde{\nabla}_if_j+\tilde{\nabla}_if_j&=\partial_if_j+\partial_jf_i-\tilde{\gamma}^{km}(\partial_i\tilde{\gamma}_{mj}+\partial_j\tilde{\gamma}_{mi}-\partial_m\tilde{\gamma}_{ij})f_{k},
\label{A.12}
\end{align}
we thus obtain the compact relation
\begin{align}
\Omega^{-2}\Delta \delta g_{ij}&=-\tilde{\nabla}_if_j-\tilde{\nabla}_if_j
-2\tilde{\gamma}_{ij}T\Omega^{-1}\dot{\Omega}=-2\tilde{\nabla}_i\tilde{\nabla}_jL-\tilde{\nabla}_iL_j-\tilde{\nabla}_jL_i
-2\tilde{\gamma}_{ij}T\Omega^{-1}\dot{\Omega}.
\label{A.13}
\end{align}
Finally, given (\ref{1.8e}) we obtain
\begin{align}
-2\bar{\psi}\tilde{\gamma}_{ij} +2\tilde{\nabla}_i\tilde{\nabla}_j \bar{E} + \tilde{\nabla}_i \bar{E}_j + \tilde{\nabla}_j \bar{E}_i + 2\bar{E}_{ij}=&
-2\psi\tilde{\gamma}_{ij} +2\tilde{\nabla}_i\tilde{\nabla}_j E + \tilde{\nabla}_i E_j + \tilde{\nabla}_j E_i + 2E_{ij}
\nonumber\\
&-2\tilde{\nabla}_i\tilde{\nabla}_jL-\tilde{\nabla}_iL_j-\tilde{\nabla}_jL_i
-2\tilde{\gamma}_{ij}\Omega^{-1}\dot{\Omega}T,
\label{A.14}
\end{align}
i.e.,
\begin{align}
2\tilde{\gamma}_{ij}(\bar{\psi}-\psi-\Omega^{-1}\dot{\Omega}T)
-2\tilde{\nabla}_i\tilde{\nabla}_j(\bar{E}-E+L)
-\tilde{\nabla}_i(\bar{E}_j-E_j+L_j)-\tilde{\nabla}_j(\bar{E}_i-E_i+L_i)-2\bar{E}_{ij}+2E_{ij}=0.
\label{A.15}
\end{align}

To extract out the required gauge invariant combinations we need to isolate the various contributions in (\ref{A.15}). To simplify the writing we rewrite (\ref{A.15}) generically as 
\begin{align}
S\tilde{\gamma}_{ij}
-\tilde{\nabla}_i\tilde{\nabla}_jQ
-\tilde{\nabla}_iA_j-\tilde{\nabla}_jA_i-R_{ij}=0.
\label{A.16}
\end{align}
Since the spatial sector is maximally three-symmetric with a Riemann tensor that is given in (\ref{1.5e}), we can use the associated geometric identities that we had used in \cite{Phelps2019} and \cite{Mannheim2020} to treat an analogous geometric situation. Specifically, we showed that for any three-scalar $S$ we have
\begin{eqnarray}
\tilde{\nabla}_a\tilde{\nabla}^a\tilde{\nabla}_iS&=&\tilde{\nabla}_i\tilde{\nabla}_a\tilde{\nabla}^aS+2k\tilde{\nabla}_iS,\quad \tilde{\nabla}_a\tilde{\nabla}^a\tilde{\nabla}_i\tilde{\nabla}_jS=\tilde{\nabla}_i\tilde{\nabla}_j\tilde{\nabla}_a\tilde{\nabla}^aS+(6k\tilde{\nabla}_i\tilde{\nabla}_j-2k\tilde{\gamma}_{ij}\tilde{\nabla}_a\tilde{\nabla}^a)S,
\nonumber\\
\tilde{\nabla}_a\tilde{\nabla}_b\tilde{\nabla}_i\tilde{\nabla}_jS&=&
\tilde{\nabla}_i\tilde{\nabla}_j \tilde{\nabla}_a\tilde{\nabla}_bS+2k\tilde{\gamma}_{ab}\tilde{\nabla}_i\tilde{\nabla}_jS
-2k\tilde{\gamma}_{ij}\tilde{\nabla}_a\tilde{\nabla}_bS
+k\tilde{\gamma}_{aj}\tilde{\nabla}_b\tilde{\nabla}_iS-k\tilde{\gamma}_{bi}\tilde{\nabla}_a\tilde{\nabla}_jS.
\label{A.17}
\end{eqnarray}
Similarly for any three-vector $A_i$ we have 
\begin{eqnarray}
&&\tilde\nabla_i\tilde\nabla_a\tilde\nabla^aA_j-\tilde\nabla_a\tilde\nabla^a\tilde\nabla_iA_j 
= 2k\tilde{\gamma}_{ij}\tilde{\nabla}_aA^a-2k(\tilde\nabla_i A_j + \tilde\nabla_j A_i),
\nonumber\\
&&\tilde\nabla^j\tilde\nabla_a\tilde\nabla^aA_j=
(\tilde\nabla_a\tilde\nabla^a+2k)\tilde\nabla^j A_j,\quad \tilde{\nabla}^j\tilde{\nabla}_iA_j=\tilde{\nabla}_i\tilde{\nabla}^jA_j+2kA_i.
\label{A.18}
\end{eqnarray}
We now apply $\tilde{\nabla}^i\tilde{\nabla}^j$ to the vector sector and obtain
\begin{align}
\tilde{\nabla}^i\tilde{\nabla}^j(\tilde{\nabla}_iA_j+\tilde{\nabla}_jA_i)=\tilde{\nabla}^i(\tilde{\nabla}_i\tilde{\nabla}^jA_j+2kA_i)
+(\tilde\nabla_a\tilde\nabla^a+6k)\tilde{\nabla}^iA_i-4k\tilde{\nabla}^iA_i.
\label{A.19}
\end{align}
Since the $A_i$ of interest obeys $\tilde{\nabla}_iA^i=0$, the right-hand side of (\ref{A.19}) vanishes. And since $\tilde{\nabla}^i\tilde{\nabla}^jR_{ij}=0$ for the $R_{ij}$ of interest to us, from (\ref{A.16}) we obtain
\begin{align}
\tilde{\nabla}^i\tilde{\nabla}^j[S\tilde{\gamma}_{ij}
-\tilde{\nabla}_i\tilde{\nabla}_jQ]=0,
\label{A.20}
\end{align}
i.e.,
\begin{align}
\tilde{\nabla}_i\tilde{\nabla}^iS-\tilde{\nabla}^i\tilde{\nabla}^j\tilde{\nabla}_j\tilde{\nabla}_iQ=0.
\label{A.21}
\end{align}
Using (\ref{A.17}) we thus obtain
\begin{align}
\tilde{\nabla}^2[S-\tilde{\nabla}^2Q-2kQ]=0.
\label{A.22}
\end{align}
Applying $\tilde{\gamma}^{ij}$ to (\ref{A.16}) we obtain
\begin{align}
3S-\tilde{\nabla}^2Q=0.
\label{A.23}
\end{align}
Consequently we obtain
\begin{align}
(\tilde{\nabla}^2+3k)S=0,\quad \tilde{\nabla}^2(\tilde{\nabla}^2+3k)Q=0.
\label{A.24}
\end{align}
These equations fall into the generic class of equations of the form $(\tilde{\nabla}^2+A_S)S=0$ discussed above, where now $A_S=0$ or $A_S=3k$. For both of these cases the only allowed solutions with $k<0$ that vanish at infinity and are well behaved at the origin have to have $S=0$  \cite{Phelps2019}. Thus under these boundary conditions we establish that in (\ref{A.16}) both $S$ and $Q$ are zero. 

Applying $\tilde{\nabla}^i$ to the remainder of (\ref{A.16}) we now obtain
\begin{align}
(\tilde{\nabla}_j\tilde{\nabla}^j+2k)A_i=0.
\label{A.25}
\end{align}
This equation falls into the generic class of equations of the form $(\tilde{\nabla}^2+A_V)A_i=0$  where $A_i$ is a three-vector and $A_V$ is a constant. However, it was shown in \cite{Phelps2019} that for $k<0$ and $A_V=2k$ the only allowed solutions that vanish at infinity and are well behaved at the origin have to have $A_i=0$. Thus in (\ref{A.25}) we set $A_i=0$. Returning to (\ref{A.16}) all that now remains is $R_{ij}$, and so it must be zero too. Thus on comparing (\ref{A.15}) with (\ref{A.16}) we obtain our sought after relations. Thus for the entire $\Delta \delta g_{\mu\nu}$  we obtain
\begin{align}
\bar{\phi}&=\phi-\dot{T} - \Omega^{-1}\dot{\Omega}T,\quad \bar{B}=B+T-\dot{L},\quad \bar{B}_i=B_i-\dot{L}_i, 
\nonumber\\
\bar{\psi}&=\psi+\Omega^{-1}\dot{\Omega}T,\quad \bar{E}=E-L,\quad \bar{E}_i=E_j-L_i,\quad \bar{E}_{ij}=E_{ij}.
\label{A.26}
\end{align}
Consequently, in the $k<0$ conformal time case the full set of gauge invariants in the metric sector is
\begin{align}
&\alpha=\phi + \psi + \frac{\partial B}{\partial\tau} - \frac{\partial^2E}{\partial\tau^2},\quad  \gamma= - \left(\frac{d \Omega}{d \tau}\right)^{-1}\Omega \psi + B - \frac{\partial E}{\partial\tau},\quad  B_i-\frac{\partial E_i}{\partial \tau},  \quad E_{ij}.
\label{A.27}
\end{align}
With the gauge transformations associated with $\Delta \delta T_{\mu\nu}$ that were given in Sec. \ref{S4} holding for all values of $k$, in the $k<0$ conformal time case the following combinations are gauge invariant:
\begin{align}
&\delta\hat{\rho}=\delta \rho -3(\rho+p)\psi,\quad X_i,\quad \hat{X}=X-\left(\frac{d \Omega}{d \tau}\right)^{-1}\Omega\psi,\quad \beta=B-\frac{\partial E}{\partial \tau}-X,
\quad
\delta \hat{p}=\delta p+\Omega\left(\frac{d\Omega}{d\tau }\right)^{-1}\frac{dp}{d\tau}\psi.
\label{A.28}
\end{align}
Strikingly, as we see,  in neither (\ref{A.27}) or (\ref{A.28}) is there any dependence on $k$. We can thus anticipate that these very same gauge combinations will apply when $k>0$ as well. Below we shall confirm this expectation by studying fluctuations in Einstein gravity, and will show that in the arbitrary $k$ case it is  precisely these particular $k$-independent combinations that appear in $\delta G_{\mu\nu}+8\pi G T_{\mu\nu}$ even as   $\delta G_{\mu\nu}+8\pi G T_{\mu\nu}$ itself explicitly depends on $k$. This lack of dependence of the gauge invariant combinations on $k$ can be understood as follows. As shown in Sec. \ref{S7} we can write a general $k$-dependent Robertson-Walker metric in the conformal to flat form given in (\ref{7.16h}) or (\ref{7.21h}) in which the only dependence on $k$ is in the conformal factor. Moreover, we also showed that the temperature fluctuations on the light cone are not dependent on the conformal factor, and thus they are not dependent on $k$. However, the temperature fluctuations on the light cone are gauge invariant. Thus the gauge invariant combinations that appear in the temperature fluctuations cannot depend on $k$ either, just as we have found. 
\section{More on the Gauge Structure of the Fluctuation Energy-Momentum Tensor}
\label{SB}
In (\ref{4.18g}) we derived an expression for $\Delta \delta T_{ij}$ of the form
\begin{align}
\Delta \delta T_{ij}&=-\Omega^2p\tilde{\gamma}_{ik}(f_m\partial_j\tilde{\gamma}^{km}+\tilde{\gamma}^{km}\partial_jf_m)
-\Omega^2p\tilde{\gamma}_{jk}(f_m\partial_i\tilde{\gamma}^{km}+\tilde{\gamma}^{km}\partial_if_m)
\nonumber\\
&-\tilde{\gamma}_{ij}T(\Omega^2\dot{p}+2\Omega\dot{\Omega}p)-\tilde{\gamma}^{km}\Omega^2p f_m\partial_k\tilde{\gamma}_{ij},
\label{B.1}
\end{align}
where $f_m=L_m+\tilde{\nabla}_mL$.  In Sec. \ref{S4} we only studied the trace of $\Delta \delta T_{ij}$ since that provided all that we needed. However, it is still of interest to study the implications of $\Delta \delta T_{ij}$ itself, and to this end we  utilize the results that we have just obtained for $\Delta \delta g_{ij}$. Inspection of (\ref{A.10}) shows  that $\Delta \delta T_{ij}$ and $\Delta \delta g_{ij}$ are very similar, as of course must be the case since (\ref{4.3g}) and (\ref{A.1}) both involve the gauge transformation of a rank two tensor. On comparing (\ref{4.3g}) and (\ref{A.1}), then given (\ref{A.13}) we can thus write $\Delta \delta T_{ij}$ as 
\begin{align}
\Omega^{-2}\Delta \delta T_{ij}&=-2p\tilde{\nabla}_i\tilde{\nabla}_jL-p\tilde{\nabla}_iL_j-p\tilde{\nabla}_jL_i
-2p\tilde{\gamma}_{ij}T\Omega^{-1}\dot{\Omega}-\dot{p}\tilde{\gamma}_{ij}T.
\label{B.2}
\end{align}
Then with $\Omega^{-2}\delta T_{ij}=pf_{ij}+\delta p \tilde{\gamma}_{ij}$ we obtain
\begin{align}
p[-2\bar{\psi}\tilde{\gamma}_{ij} +2\tilde{\nabla}_i\tilde{\nabla}_j \bar{E} + \tilde{\nabla}_i \bar{E}_j + \tilde{\nabla}_j \bar{E}_i + 2\bar{E}_{ij}]+\delta \bar{p} \tilde{\gamma}_{ij}&=
p[-2\psi\tilde{\gamma}_{ij} +2\tilde{\nabla}_i\tilde{\nabla}_j E + \tilde{\nabla}_i E_j + \tilde{\nabla}_j E_i + 2E_{ij}]+\delta p \tilde{\gamma}_{ij}
\nonumber\\
&+p[-2\tilde{\nabla}_i\tilde{\nabla}_jL-\tilde{\nabla}_iL_j-\tilde{\nabla}_jL_i
-2\tilde{\gamma}_{ij}\Omega^{-1}\dot{\Omega}T]-\dot{p}\tilde{\gamma}_{ij}T.
\label{B.3}
\end{align}
Thus with the same boundary conditions that we used for $\Delta \delta g_{ij}$ we identify 
\begin{align}
-2p\bar{\psi}+\delta \bar{p}=-2p\psi+\delta p-2p\Omega^{-1}\dot{\Omega}T-\dot{p}T,\quad  \bar{E}=E-L,\quad \bar{E}_i=E_i-L_i, \quad \bar{E}_{ij}=E_{ij}.
\label{B.4}
\end{align}
We thus recover the  $\bar{E}=E-L$, $\bar{E}_i=E_i-L_i$, $\bar{E}_{ij}=E_{ij}$ relations that we obtained in (\ref{A.26}) from a study of $\Delta \delta g_{ij}$. And with the $\bar{\psi}=\psi+\Omega^{-1}\dot{\Omega}T$ relation that is also given in (\ref{A.26}) we obtain $\delta \bar{p}=\delta p-\dot{p}T$, just as given  (\ref{4.24g}). Thus the only information in $\Delta \delta T_{ij}$ that is not also contained in $\Delta \delta g_{ij}$ is contained in $\tilde{\gamma}^{ij}\Delta \delta T_{ij}$. Given our analysis of $\Delta \delta T_{ij}$ and $\Delta \delta g_{ij}$ in the non-zero $k$ case we thus confirm that all the gauge invariant combinations that we have discussed in this paper apply in the arbitrary $k$ case even as they do not explicitly  depend on $k$ at all.

\section{Gauge Invariance of the Fluctuation Equations}
\label{SC}

To buttress our analysis of gauge invariance we evaluate the Einstein gravity $\Delta_{\mu\nu}=\delta G_{\mu\nu}+8\pi G \delta T_{\mu\nu}$ for fluctuations around a $k\neq 0$ Robertson-Walker background in which $G_{\mu\nu}+8\pi G T_{\mu\nu}=0$. For the conformal time metric with $k\neq 0$ given in (\ref{1.8e}), on setting $8\pi G=1$ and using the definition of $\hat{X}$ and $X_i$ given in this paper  the various components of the conformal $\Delta_{\mu\nu}$ take the form \cite{Phelps2019}
\begin{align}
\Delta_{\tau\tau}&= 6 \dot{\Omega}^2 \Omega^{-2}(\alpha-\dot\gamma) + \delta \hat{\rho} \Omega^2 + 2 \dot{\Omega} \Omega^{-1} \tilde{\nabla}_{a}\tilde{\nabla}^{a}\gamma, 
\nonumber\\
\Delta_{\tau i}&= -2 \dot{\Omega} \Omega^{-1} \tilde{\nabla}_{i}(\alpha - \dot\gamma) + 2 k \tilde{\nabla}_{i}\gamma 
+(-4 \dot{\Omega}^2 \Omega^{-2}  + 2 \overset{..}{\Omega} \Omega^{-1}  - 2 k) \tilde{\nabla}_{i}\hat{X}
\nonumber\\
& +k(B_i-\dot E_i)+ \frac{1}{2} \tilde{\nabla}_{a}\tilde{\nabla}^{a}(B_{i} - \dot{E}_{i})
+ (-4 \dot{\Omega}^2 \Omega^{-2} + 2 \overset{..}{\Omega} \Omega^{-1} - 2 k )X_{i},
\nonumber\\
\Delta_{ij}&= \tilde{\gamma}_{ij}\big[ 2 \dot{\Omega}^2 \Omega^{-2}(\alpha-\dot\gamma)
-2  \dot{\Omega} \Omega^{-1}(\dot\alpha -\ddot\gamma)-4\ddot\Omega\Omega^{-1}(\alpha-\dot\gamma)+ \Omega^2 \delta \hat{p}-\tilde\nabla_a\tilde\nabla^a( \alpha + 2\dot\Omega \Omega^{-1}\gamma) \big] +\tilde\nabla_i\tilde\nabla_j( \alpha + 2\dot\Omega \Omega^{-1}\gamma)
\nonumber\\
&
+\dot{\Omega} \Omega^{-1} \tilde{\nabla}_{i}(B_{j}-\dot E_j)+\frac{1}{2} \tilde{\nabla}_{i}(\dot{B}_{j}-\ddot{E}_j)
+\dot{\Omega} \Omega^{-1} \tilde{\nabla}_{j}(B_{i}-\dot E_i)+\frac{1}{2} \tilde{\nabla}_{j}(\dot{B}_{i}-\ddot{E}_i)
\nonumber\\
&- \overset{..}{E}_{ij} - 2 k E_{ij} - 2 \dot{E}_{ij} \dot{\Omega} \Omega^{-1} + \tilde{\nabla}_{a}\tilde{\nabla}^{a}E_{ij}.
\label{C.1}
\end{align}
Here the dot denotes $d/d\tau$ and the quantities that appear in the conformal time $\Delta_{\mu\nu}$ are precisely the ones that appear in (\ref{4.25g}) \cite{footnoteK}. Since these are the only quantities that appear, we see that the conformal time $\Delta_{\mu\nu}$ is built entirely out of gauge invariant combinations, just as it should be.

Analogously, in comoving time we have 
\begin{align}
\Delta_{tt}=& \frac{1}{a^2}\bigg{[}6 \dot{a}^2(\alpha- a\dot{\gamma})+\delta\hat{\rho} a ^2+2 \dot{a} \tilde{\nabla}_{a}\tilde{\nabla}^{a}\gamma\bigg{]},
\nonumber\\
\Delta_{ti}=&\frac{1}{a}\bigg{[}-2 \dot{a} \tilde{\nabla}_{i}\left(\alpha -a  \dot{\gamma}\right)+2 k \tilde{\nabla}_{i}\gamma 
+ \left[-4 \dot{a}^2+2 \left(\dot{a}^2+a  \ddot{a}\right)-2 k\right]\tilde{\nabla}_{i}\hat{X}
\nonumber\\
&+k \left(B_i-a  \dot{E}_i\right)+\frac{1}{2} \tilde{\nabla}_{a}\tilde{\nabla}^{a}(B_i-a  \dot{E}_i)+ \left[-4 \dot{a}^2
+2 \left(\dot{a}^2+a  \ddot{a}\right)-2 k\right]X_i\bigg{]},
\nonumber\\
\Delta_{ij}=&\tilde{\gamma}_{ij}\Big[2 \dot{a}^2 \left(\alpha -a  \dot{\gamma}\right) -2 \dot{a} \left(a  \dot{\alpha}-a\dot{a}\dot{\gamma} -a^2  \ddot{\gamma}\right)-4 \left(\dot{a}^2+a  \ddot{a}\right) \left(\alpha -a  \dot{\gamma}\right)
+\delta\hat{p} a^2
-\tilde{\nabla}_a\tilde{\nabla}^a \left(\alpha +2\dot{a}  \gamma\right)\Big]+\tilde{\nabla}_i\tilde{\nabla}_j(\alpha +2 \dot{a}\gamma)
\nonumber\\
&+\dot{a} \tilde{\nabla}_{i}\left(B_j-a  \dot{E}_j\right)
+\frac{1}{2}\tilde{\nabla}_i(a \dot{B}_j-a\dot{a} \dot{E}_j-a^2  \ddot{E}_j)
+\dot{a} \tilde{\nabla}_{j}\left(B_i-a  \dot{E}_i\right)
+\frac{1}{2}\tilde{\nabla}_j(a  \dot{B}_i-a\dot{a} \dot{E}_i-a^2  \ddot{E}_i)
\nonumber\\
&-a^2 \ddot{E}_{ij} -a  \dot{a} \dot{E}_{ij}-2 k E_{ij}- 2a  \dot{a} \dot{E}_{ij}+\tilde{\nabla}_{a}\tilde{\nabla}^{a}E_{ij}.
\label{C.2}
\end{align}
Here the dot denotes $d/dt$ and the quantities that appear in the comoving time $\Delta_{\mu\nu}$ are precisely the ones that appear in (\ref{4.26g}). Since these are the only quantities that appear, we see that the comoving time $\Delta_{\mu\nu}$ is built entirely out of gauge invariant combinations, just as it should be.

To obtain further insight into the relation of the gauge invariant combinations to the conformal structure of light cone fluctuations that we have found in this paper, it is instructive to consider a gravitational theory that it is itself conformal invariant. Einstein gravity is not conformal, which is why even the conformal time $\Delta_{\mu\nu}$ given in (\ref{C.1}) depends on gauge invariants such as the $\Omega$-dependent quantity $\gamma$ even as such $\Omega$-dependent quantities  do not appear in the light cone temperature fluctuation. However, the conformal gravity theory (see e.g. the reviews in \cite{Mannheim2006,Mannheim2012b,Mannheim2017}) is conformal invariant. Conformal gravity is based on an action that is invariant under $g_{\mu\nu}(x)\rightarrow e^{2\alpha(x)}g_{\mu\nu}(x)$, viz. 
\begin{eqnarray}
I_{\rm W}=-\alpha_g\int d^4x\, (-g)^{1/2}C_{\lambda\mu\nu\kappa}
C^{\lambda\mu\nu\kappa}
\equiv -2\alpha_g\int d^4x\, (-g)^{1/2}\left[R_{\mu\kappa}R^{\mu\kappa}-\frac{1}{3} (R^{\alpha}_{\phantom{\alpha}\alpha})^2\right].
\label{C.3}
\end{eqnarray}
Here $\alpha_g$ is a dimensionless  gravitational coupling constant, and $C_{\lambda\mu\nu\kappa}$ is the conformal Weyl tensor. Functional variation with respect to the metric $g_{\mu\nu}(x)$ generates fourth-order derivative gravitational equations of motion of the form (see e.g. \cite{Mannheim2006})
\begin{eqnarray}
-\frac{2}{(-g)^{1/2}}\frac{\delta I_{\rm W}}{\delta g_{\mu\nu}}=4\alpha_g W^{\mu\nu}=4\alpha_g\left[2\nabla_{\kappa}\nabla_{\lambda}C^{\mu\lambda\nu\kappa}-
R_{\kappa\lambda}C^{\mu\lambda\nu\kappa}\right]=4\alpha_g\left[W^{\mu
\nu}_{(2)}-\frac{1}{3}W^{\mu\nu}_{(1)}\right]=T^{\mu\nu},
\label{C.4}
\end{eqnarray}
where the functions $W^{\mu \nu}_{(1)}$ and $W^{\mu \nu}_{(2)}$ (respectively associated with the $(R^{\alpha}_{\phantom{\alpha}\alpha})^2$ and $R_{\mu\kappa}R^{\mu\kappa}$ terms in (\ref{C.3})) are given by
\begin{eqnarray}
W^{\mu \nu}_{(1)}&=&
2g^{\mu\nu}\nabla_{\beta}\nabla^{\beta}R^{\alpha}_{\phantom{\alpha}\alpha}                                             
-2\nabla^{\nu}\nabla^{\mu}R^{\alpha}_{\phantom{\alpha}\alpha}                          
-2 R^{\alpha}_{\phantom{\alpha}\alpha}R^{\mu\nu}                              
+\frac{1}{2}g^{\mu\nu}(R^{\alpha}_{\phantom{\alpha}\alpha})^2,
\nonumber\\
W^{\mu \nu}_{(2)}&=&
\frac{1}{2}g^{\mu\nu}\nabla_{\beta}\nabla^{\beta}R^{\alpha}_{\phantom{\alpha}\alpha}
+\nabla_{\beta}\nabla^{\beta}R^{\mu\nu}                    
 -\nabla_{\beta}\nabla^{\nu}R^{\mu\beta}                       
-\nabla_{\beta}\nabla^{\mu}R^{\nu \beta}                          
 - 2R^{\mu\beta}R^{\nu}_{\phantom{\nu}\beta}                                    
+\frac{1}{2}g^{\mu\nu}R_{\alpha\beta}R^{\alpha\beta}.
\label{C.5}
\end{eqnarray}                                 
The fluctuation $\delta W_{\mu\nu}$ in the conformal time $W_{\mu\nu}$ is of the form \cite{Phelps2019}
\begin{eqnarray}
\delta W_{\tau\tau}&=& - \frac{2}{3\Omega^2} (\tilde\nabla_a\tilde\nabla^a + 3k)\tilde\nabla_b\tilde\nabla^b \alpha,
 \nonumber\\ 
\delta W_{\tau i}&=& -\frac{2}{3\Omega^2}  \tilde\nabla_i (\tilde\nabla_a\tilde\nabla^a + 3k)\dot\alpha
+\frac{1}{2\Omega^2}(\tilde\nabla_b \tilde\nabla^b-\partial_{\tau}^2-2k)(\tilde\nabla_c \tilde\nabla^c+2k)(B_i-\dot{E}_i),
  \nonumber\\ 
\delta W_{ij}&=& -\frac{1}{3 \Omega^2} \left[ \tilde{\gamma}_{ij} \tilde\nabla_a\tilde\nabla^a (\tilde\nabla_b \tilde\nabla^b +2k-\partial_{\tau}^2)\alpha - \tilde\nabla_i\tilde\nabla_j(\tilde\nabla_a\tilde\nabla^a - 3\partial_{\tau}^2)\alpha \right]
\nonumber\\
&& +\frac{1}{2 \Omega^2} \left[ \tilde\nabla_i (\tilde\nabla_a\tilde\nabla^a -2k-\partial_{\tau}^2) (\dot{B}_j-\ddot{E}_j) 
+  \tilde\nabla_j ( \tilde\nabla_a\tilde\nabla^a -2k-\partial_{\tau}^2) (\dot{B}_i-\ddot{E}_i)\right]
\nonumber\\
&&+ \frac{1}{\Omega^2}\left[ (\tilde\nabla_b \tilde\nabla^b-\partial_{\tau}^2-2k)^2+4k\partial_{\tau}^2 \right] E_{ij}.
\label{C.6}
\end{eqnarray}
As we see, only the $\Omega(\tau)$-independent gauge invariant metric combinations $\alpha$, $B_i-\dot{E}_i$ and $E_{ij}$ appear in the conformal-gravity based $\delta W_{\mu\nu}$, with the $\Omega(\tau)$-dependent $\gamma$ being excluded. Consequently, conformal invariance excludes the presence of any gauge invariant combination  that has any  dependence on the conformal factor  $\Omega(\tau)$, just as we found with  photon temperature fluctuations. Since for these temperature fluctuations any dependence on the three-curvature $k$ is also excluded, this is a quite remarkable  occurrence.


\begin{thebibliography}{99}
\bibitem{footnoteA} If any specific gauge is chosen one cannot make large gauge transformations to it as they will cause the fluctuations to diverge at infinity. In a gauge invariant formulation  a fluctuation that is bounded in one gauge is bounded in every gauge.
\bibitem{footnoteD} In making the transformation $d\tau=dt/a(t)$, $\Omega(\tau)=a(t)$,  an expression such  as $-2a(t)(\tilde{\nabla}_iB+B_i)dtdx_i$ transforms into  $-2\Omega^2(\tau)(\tilde{\nabla}_iB+B_i)d\tau dx_i$. The various $a(t)$ and $\Omega(\tau)$ factors that respectively appear in (\ref{1.2e}) and (\ref{1.8e}) have been chosen so that the SVT components themselves transform into each other, so that, for instance, $B(t,x_i)\rightarrow B(\tau,x_i)$.
\bibitem{footnoteE} We set $dr/dt=-1/a(t)$ rather than $dr/dt=+1/a(t)$ because  we set $r=0$ at the current late time rather than at some early time, with $r$ decreasing as $t$ increases.
\bibitem{footnoteF} To lowest order in the perturbation we can set $r(t)=s(t)$ in $N[r(t)\hat{n},t]$.
\bibitem{footnoteG} Technically it should be the $s(t)$ that leaves $r_L+\delta t_L\delta u^r(r_L\hat{n},t_L)$ at $t_L+\delta t_L$, but that would produce a higher-order correction.
\bibitem{footnoteH} A lightlike radial background photon mode has a velocity four-vector  of the form $dx^{\mu}/d\lambda=(dt/d\lambda, dr/d \lambda,0,0)$, where $\lambda$ is an affine parameter that parameterizes the trajectory. Since the mode is on the light cone $\lambda$ cannot be the proper time. However, we could take it to be $t$, in which case the four-vector takes the form $dx^{\mu}/dt=(1,dr/dt,0,0)$. In a $k=0$ Robertson-Walker geometry the lightlike  radial background photon four-vector is given by $dx^{\mu}/dt=(1,-1/a,0,0)$. In contrast, the fluid velocity four-vector (for any spatial three-curvature) is given by $u^{\mu}=(1,0,0,0)$.
\bibitem{footnoteI} Even if some general vector $V^i$ is transverse it can still have a component $\tilde{\nabla}^iV$ that takes the form of the derivative of a scalar $V$  if the scalar obeys $\tilde{\nabla}_i\tilde{\nabla}^iV=0$.  Thus even when written as the derivative of a scalar such a component is still transverse. A longitudinal component can also be written as the derivative of a scalar $Y$. However such a scalar would not obey $\tilde{\nabla}_i\tilde{\nabla}^iY=0$.
\bibitem{footnoteL} While these gauge invariant combinations have no dependence on $k$, once we introduce a dynamical evolution equation and solve for an explicit form for them, the form that we will obtain will depend on $k$ since the dynamical evolution equations do.
\bibitem{footnoteM} The inclusion of the $\delta T(0,t_0)/T_0$ term is of little practical consequence since, as noted in \cite{Weinberg2008},  the entire late term only contributes to the $\ell=0$ and $\ell=1$ angular momentum components of the temperature fluctuations, and not to the ones with $\ell \geq 2$ that are critical to the understanding of the anisotropy of the CMB.
\end{thebibliography}
\end{document}